\documentclass{revtex4}%
\usepackage{amsmath}
\usepackage{amsfonts}
\usepackage{amssymb}
\usepackage{graphicx}%
\providecommand{\U}[1]{\protect\rule{.1in}{.1in}}
\begin{document}
\title{The explicit expression of the fugacity for weakly interacting Bose and Fermi gases}
\author{Wu-Sheng Dai}
\email{daiwusheng@tju.edu.cn}
\author{Mi Xie }
\email{xiemi@tju.edu.cn}
\affiliation{Department of Physics, Tianjin University, Tianjin 300072, P. R. China }
\affiliation{LiuHui Center for Applied Mathematics, Nankai University \& Tianjin
University, Tianjin 300072, P. R. China}
\date{}

\begin{abstract}
In this paper, we calculate the explicit expression for the fugacity for two-
and three-dimensional weakly interacting Bose and Fermi gases from their
equations of state in isochoric and isobaric processes, respectively, based on
the mathematical result of the boundary problem of analytic functions --- the
homogeneous Riemann-Hilbert problem. We also discuss the Bose-Einstein
condensation phase transition of three-dimensional hard-sphere Bose gases.

\end{abstract}
\maketitle

%\noindent PACS codes: 05.30.-d, 05.70.Ce, 02.30.Rz

%\noindent Keywords: hard-sphere quantum gas, fugacity, Riemann-Hilbert problem

\section{Introduction \label{I}}

\textit{The physical problem.} The equation of state for a quantum gas can be
obtained by\ eliminating the fugacity $z$ between the equations%
\begin{align}
\frac{P}{kT}  &  =\frac{1}{V}\ln\Xi\left(  T,z,V\right)  ,\label{I1}\\
n  &  =\frac{1}{V}z\frac{\partial}{\partial z}\ln\Xi\left(  T,z,V\right)  ,
\label{I2}%
\end{align}
where $P$ is the pressure, $\ln\Xi$ the grand potential, $n$ the particle
number density, and $V$ the volume. Concretely, for example, in isochoric
processes, solving $z=z\left(  n,T\right)  $ from Eq. (\ref{I2}) and
substituting it into Eq. (\ref{I1}) give the equation of state
\begin{equation}
\frac{P}{kT}=\left.  \frac{1}{V}\ln\Xi\left(  T,z,V\right)  \right\vert
_{z=z\left(  n,T\right)  },
\end{equation}
and in isobaric processes, solving $z=z\left(  P,T\right)  $ from Eq.
(\ref{I1}) and substituting it into Eq. (\ref{I2}) give the equation of state
\begin{equation}
n=\left.  \frac{1}{V}z\frac{\partial}{\partial z}\ln\Xi\left(  T,z,V\right)
\right\vert _{z=z\left(  P,T\right)  }.
\end{equation}
In a word, to solve, e.g., the pressure $P$ requires to solve the fugacity $z
$ first. However, only in some simple cases, such as classical ideal gases
\cite{Kadanoff}, two-dimensional ideal quantum gases, the explicit expression
of the fugacity $z$ can be obtained by solving Eq. (\ref{I2}) or (\ref{I1}).
In most cases, the fugacity can only be obtained approximately, e.g., at
high-temperature and low-density or low-temperature and high-density limit. In
this paper, we will calculate the explicit expression for the fugacity for
two- and three-dimensional weakly interacting hard-sphere Bose and Fermi gases
by directly solving Eqs. (\ref{I2}) and (\ref{I1}) for isochoric and isobaric
processes, respectively.

The hard-sphere gas, as a simplified model, is of great value for
investigating the more general theory of interacting gases and can be extended
to some more general cases. This is because a particle that is spread out in
space sees only an averaged effect of the potential and, thus, often a
complete knowledge of the detailed interaction potential is not necessary for
a satisfactory description \cite{HYL,HYL2}. Or, from the viewpoint of quantum
mechanics, due to the low collision energy of collisions among the gas
molecules, the shape-independent $s$-wave contribution dominates. The
equations of state for three-dimensional hard-sphere Bose and Fermi gases are
presented in Ref. \cite{LeeYang} by the binary collision expansion method. The
equations of state for two-dimensional cases is discussed in Ref.
\cite{OurEPL}. For one-dimensional cases, there are deeply analyses in Ref.
\cite{Yang,Yang2,Yang3,Yang4,Yang5,Yang6,Yang7,Yang8,Yang9}. Non-ideal quantum
gases are studied from various aspects, e.g., equivalence between hard-sphere
and zero-range potentials \cite{Val}, thermodynamics \cite{RD,HLD},
ferromagnetic transitions \cite{SMB,HH}, and the upper limit on $T_{c}$ of BEC
\cite{OurAnn}. The second and third virial coefficients for strongly
correlated fermions are discussed for two- \cite{LHD10} and three-dimensional
\cite{LHD09} cases by using the few-particle exact solutions. Many methods are
employed, e.g., auxiliary field \cite{MDC} and cluster expansion \cite{SKU} methods.

\textit{The mathematical method. }The key idea of the method for solving the
explicit expression for the fugacity from Eqs. (\ref{I2}) and (\ref{I1}) is
based on the boundary problem of analytic functions --- the homogeneous
Riemann-Hilbert problem \cite{Muskhelishvili}. In a Riemann-Hilbert problem,
one seeks to find a sectionally analytic function $\psi\left(  \zeta\right)  $
under the boundary condition%
\begin{equation}
\psi^{+}\left(  \zeta\right)  =G\left(  \zeta\right)  \psi^{-}\left(
\zeta\right)  +g\left(  \zeta\right)  ,\,\,\zeta\in L,
\end{equation}
where $L$ is a union of a finite number of smooth simple arcs, $\psi
^{+}\left(  \zeta\right)  $ and $\psi^{-}\left(  \zeta\right)  $ are boundary
values of $\psi\left(  \zeta\right)  $ on the left and right of $L$, $G\left(
\zeta\right)  $ and $g\left(  \zeta\right)  $ are functions satisfying
$\left\vert f\left(  \zeta_{2}\right)  -f\left(  \zeta_{1}\right)  \right\vert
\leq A\left\vert \zeta_{2}-\zeta_{1}\right\vert ^{\mu}$ with $A$, $\mu>0$, the
H\"{o}lder condition, and $G\left(  \zeta\right)  \neq0$ everywhere on $L$.
The function $\psi\left(  \zeta\right)  $ being a sectionally analytic
function means that $\psi\left(  \zeta\right)  $ is analytic in each region
not containing points of the boundary $L$ and is continuous on $L$ from the
left and from the right, excepting possibly some ends of $L$ and near such
ends the function $\psi\left(  \zeta\right)  $ should satisfy
\begin{equation}
\left\vert \psi\left(  \zeta\right)  \right\vert <\frac{constant}{\left\vert
\zeta-c_{k}\right\vert ^{\alpha_{k}}},\,\,\,\alpha_{k}\,<1,
\end{equation}
where $\alpha_{k}$ is a constant corresponding to the $k$-th end $c_{k}$. The
homogeneous Riemann-Hilbert problem is a Riemann-Hilbert problem with
$g\left(  \zeta\right)  =0$. In a homogeneous Riemann-Hilbert problem, the
problem is converted into finding a function $\psi\left(  \zeta\right)  $ from
the jump on the two sides of the boundary $L$, $G\left(  \zeta\right)
=\psi^{+}\left(  \zeta\right)  /\psi^{-}\left(  \zeta\right)  \,$.

Take an isochoric process as an example. In this case, the fugacity is
determined by Eq. (\ref{I2}). The first step is to introduce a real function
whose zero gives Eq. (\ref{I2}). Analytically continuing this real function to
the complex plane gives a complex function, and, clearly, the zero of the real
function becomes the zero of this complex function on the real axis.
Therefore, the fugacity $z$ becomes a real zero of this complex function, and
the problem of solving the fugacity is converted into a problem of seeking the
real zero of such a complex function. The key step toward solving the zero
relies on the homogeneous Riemann-Hilbert problem.

Based on the Riemann-Hilbert problem, Leonard calculates the explicit
expression for the fugacity for ideal Bose and Fermi gases \cite{Leonard}. We
also construct an exactly solvable phase transition model, the generalized
Bose--Einstein condensation, with the help of the Riemann-Hilbert problem
\cite{Ours1}.

Note that besides Bose and Fermi cases, this method can also be used to solve
the fugacity for intermediate statistics which describes various elementary
excitations \cite{OurAnn2,OurJSP} corresponding to various quantization
schemes \cite{OurAnn3}. Moreover, the method can be applied to solve the ideal
and interacting gases in confined space
\cite{JMP07,PLA,PLA2,PLA3,PLA4,PLA5,PLA6,PLA7,PLA8}.

Moreover, if a singularity appears in the relation of the fugacity $z$ and the
temperature $T$, there is a phase transition. In this paper, we give rigorous
results of the relation between $z$ and $T$, so we can judge the occurrence of
a phase transition by observing the relation of $z$ and $T$ directly.

A detailed description of the method is given in Sec. \ref{II}. In Secs.
\ref{III} and \ref{V}, we calculate the explicit expression for the fugacity
for two- and three-dimensional hard-sphere Bose and Fermi gases, isochoric
cases in Sec. \ref{III} and isobaric cases in Sec. \ref{V}. In Sec.
\ref{HTLD}, we compare our result with the virial expansion at high
temperatures and low densities. In Sec. \ref{Phasetransitions}, we discuss the
Bose-Einstein condensation phase transition of three--dimensional hard-sphere
Bose gases. The conclusions are summarized in Sec. \ref{VI}.

\section{The method\label{II}}

In this section, we give a description of the method.

\subsection{The formal explicit expression of the fugacity}

Take an isochoric process as an example for describing the method.

In an isochoric case, what we want to do is to solve an explicit expression
for the fugacity $z$ from Eq. (\ref{I2}). Introduce a real function%
\begin{equation}
\Psi\left(  z\right)  =\frac{1}{nV}z\frac{\partial}{\partial z}\ln\Xi\left(
T,z,V\right)  -1\equiv f\left(  z\right)  -1, \label{II1}%
\end{equation}
where $f\left(  z\right)  =\left(  1/nV\right)  z\left(  \partial/\partial
z\right)  \ln\Xi\left(  T,z,V\right)  $. Clearly, Eq. (\ref{I2}) corresponds
to $\Psi\left(  z\right)  =0$. In other words, the fugacity $z$ is a zero of
the function $\Psi\left(  z\right)  $, and the problem of solving the fugacity
$z$ is converted into the problem of seeking the zero of $\Psi\left(
z\right)  $.

To seek the zero of $\Psi\left(  z\right)  $, we first analytically continue
the function $\Psi\left(  z\right)  $ to the whole complex $\zeta$-plane. The
analytically continued function is
\begin{equation}
\Psi\left(  \zeta\right)  =f\left(  \zeta\right)  -1,\,\,\zeta\in\mathbb{C}.
\label{II2}%
\end{equation}
The fugacity $z$ that we want to solve is a zero of the real function
$\Psi\left(  z\right)  $ and is, of course, a zero of the complex function
$\Psi\left(  \zeta\right)  $ on the real axis. Usually, the function
$\Psi\left(  \zeta\right)  $ has not only one zero. Besides the zero
corresponding to the fugacity $z$, there are still other zeros in the complex
$\zeta$-plane, denoted as $\omega_{i}$, $\left(  i=1,2,\cdots n_{\omega
}-1\right)  $, where $n_{\omega}$ is the total number of the zeros. Moreover,
$\Psi\left(  \zeta\right)  $ may also has singularities in the complex $\zeta
$-plane. In the following, we will show that in our case, $\Psi\left(
\zeta\right)  $ has no isolated singularities, and all its singularities are
non-isolated singularities, forming some arcs in the complex $\zeta$-plane.
Such arcs, which form a boundary of the analytic region of $\Psi\left(
\zeta\right)  $, will be denoted by $L_{i}$, $\left(  i=1,2,\cdots
,m_{L}\right)  $, where $m_{L}$ is the total number of the arcs. That is to
say, $\Psi\left(  \zeta\right)  $ is analytic in each region not containing
points of the arcs $L_{i} $ and has $n_{\omega}$ zeros; at least one of the
zeros of $\Psi\left(  \zeta\right)  $ is on the real axis. Therefore, we can
express $\Psi\left(  \zeta\right)  $ in the following form:%
\begin{equation}
\Psi\left(  \zeta\right)  =\eta\left(  \zeta-z\right)
%TCIMACRO{\dprod \limits_{i=1}^{n_{\omega}-1}}%
%BeginExpansion
{\displaystyle\prod\limits_{i=1}^{n_{\omega}-1}}
%EndExpansion
\left(  \zeta-\omega_{i}\right)  \Phi\left(  \zeta\right)  , \label{II5}%
\end{equation}
where $\eta$ is a constant and $\Phi\left(  \zeta\right)  $ is a function
vanishing nowhere in the complex $\zeta$-plane and analytic in each region not
containing points of the arcs $L_{i}$. In principle, except the fugacity $z$,
if we know all other $n_{\omega}-1$ zeros of $\Phi\left(  \zeta\right)  $,
$\omega_{i}$, the explicit expression of $z$ can be formally expressed from
Eq. (\ref{II5}),%
\begin{equation}
z=\zeta-\left[  \eta%
%TCIMACRO{\dprod \limits_{i=1}^{n_{\omega}-1}}%
%BeginExpansion
{\displaystyle\prod\limits_{i=1}^{n_{\omega}-1}}
%EndExpansion
\left(  \zeta-\omega_{i}\right)  \right]  ^{-1}\frac{\Psi\left(  \zeta\right)
}{\Phi\left(  \zeta\right)  }. \label{II6}%
\end{equation}
Then, the problem of solving $z$ is converted into the problems of solving the
function $\Phi\left(  \zeta\right)  $ and finding the zeros $\omega_{i}$.

\subsection{$\Phi\left(  \zeta\right)  $ and the fundamental solution of the
homogeneous Riemann-Hilbert problem}

The function $\Phi\left(  \zeta\right)  $ can be determined with the help of
the homogeneous Riemann-Hilbert problem.

In a homogeneous Riemann-Hilbert problem, if the jump of a function on the
boundary is known, we can determine the function up to an arbitrary
polynomial. To seek the function $\Phi\left(  \zeta\right)  $, we first need
to solve the fundamental solution of the homogeneous Riemann-Hilbert problem,
denoted as $\varphi\left(  \zeta\right)  $. The fundamental solution
$\varphi\left(  \zeta\right)  $ is such a solution that $\varphi\left(
\zeta\right)  $ and its reciprocal $1/\varphi\left(  \zeta\right)  $ both are
sectionally analytic functions. More concretely, a fundamental solution has no
zeros and isolated singularities; all its singularities lie on some arcs,
forming a boundary for the analytic region; at each end of the arcs, $c_{k}$,
the degree of divergence of the solution $\varphi\left(  \zeta\right)  $ and
its reciprocal $1/\varphi\left(  \zeta\right)  $ is less than $1$, i.e.,
$\left\vert \varphi\left(  \zeta\right)  \right\vert <constant/\left\vert
\zeta-c_{k}\right\vert ^{\alpha_{k}}$ and\ $\left\vert 1/\varphi\left(
\zeta\right)  \right\vert <constant/\left\vert \zeta-c_{k}\right\vert
^{\alpha_{k}^{\prime}}$, where $\alpha_{k}$ and $\alpha_{k}^{\prime}$ are
constants less than $1$. The fundamental solution, generally speaking, is not
completely determined by the above conditions; they are divided into some
classes according to their behaviors near the ends, and any class of which can
be chosen as the fundamental solution of the homogeneous Riemann-Hilbert
problem. In this paper, we choose the fundamental solution satisfying%
\begin{equation}
\left\vert \varphi\left(  \zeta\right)  \right\vert <\frac{constant}%
{\left\vert \zeta-c_{k}\right\vert ^{\alpha_{k}}},\,\,\,0\leq\alpha_{k}\,<1.
\label{ch}%
\end{equation}

Generally speaking, the function $\Phi\left(  \zeta\right)  $ that we want to
find is not a fundamental solution though it has no zeros and isolated
singularities and all its singularities lie on the arcs, $L_{i}$
($i=1,\cdots,m_{L}$), since the degree of divergence of $\Phi\left(
\zeta\right)  $ or $1/\Phi\left(  \zeta\right)  $ at the ends of such arcs may
not be less than $1$. However, in terms of the fundamental solution
$\varphi\left(  \zeta\right)  $, the function $\Phi\left(  \zeta\right)  $ can
always be expressed as%
\begin{equation}
\Phi\left(  \zeta\right)  =\frac{\varphi\left(  \zeta\right)  }{%
%TCIMACRO{\dprod \limits_{k=1}^{n}}%
%BeginExpansion
{\displaystyle\prod\limits_{k=1}^{n}}
%EndExpansion
\left(  \zeta-c_{k}\right)  ^{\beta_{k}}}, \label{II7}%
\end{equation}
where $n$ is the number of the ends that are different from infinity and
$\beta_{k}$ is a constant determined by both the degree of divergence of
$\Phi\left(  \zeta\right)  $ and the degree of divergence of the fundamental
solution $\varphi\left(  \zeta\right)  $\ at the $k$-th end $c_{k}$. For a
chosen fundamental solution $\varphi\left(  \zeta\right)  $ (in the present
case, the fundamental solution is chosen to satisfy Eq. (\ref{ch})),
$\beta_{k}$ should ensure that the behaviors of the two sides of Eq.
(\ref{II7}) in the neighborhood of the $k$-th\ end of the arcs $L_{i}$ are the
same. For the end at infinity, we need not pay special attention since at such
an end the degree of divergence of the function encountered in the present
case is less than $1$. Concretely, near infinity, the behavior of the function
$\Psi\left(  \zeta\right)  $, defined by Eq. (\ref{II2}), lies on the
asymptotic behavior of the Bose-Einstein integral or the Fermi-Dirac integral
near infinity. Near the point of $\infty$, the Cauchy principal value of the
analytically continued Bose-Einstein integral $g_{\sigma}\left(  \zeta\right)
\sim-\left(  \ln\zeta\right)  ^{\sigma}/\left[  \sigma\Gamma\left(
\sigma\right)  \right]  $ \cite{Clunie}, and then $\left.  \Psi\left(
\zeta\right)  \right\vert _{\zeta\rightarrow\infty}\sim\left(  \ln
\zeta\right)  ^{k}$, where $k$ is a constant; near the point of $-\infty$, the
Cauchy principal value of the analytically continued Fermi-Dirac integral
$f_{\sigma}\left(  \zeta\right)  \sim\left[  \ln\left(  -\zeta\right)
\right]  ^{\sigma}/\left[  \sigma\Gamma\left(  \sigma\right)  \right]  $, and
then $\left.  \Psi\left(  \zeta\right)  \right\vert _{\zeta\rightarrow-\infty
}\sim\left[  \ln\left(  -\zeta\right)  \right]  ^{k}$.

The fundamental solution $\varphi\left(  \zeta\right)  $ can be obtained based
on the result of the homogeneous Riemann-Hilbert problem. From Eqs.
(\ref{II5}) and (\ref{II7}), we can see that the jumps of the functions
$\Psi\left(  \zeta\right)  $, $\Phi\left(  \zeta\right)  $, and $\varphi
\left(  \zeta\right)  $ at the two sides of the boundary, which consists of
arcs $L_{i}$, are the same, i.e.,%
\begin{equation}
\frac{\varphi^{+}\left(  \zeta\right)  }{\varphi^{-}\left(  \zeta\right)
}=\frac{\Phi^{+}\left(  \zeta\right)  }{\Phi^{-}\left(  \zeta\right)  }%
=\frac{\Psi^{+}\left(  \zeta\right)  }{\Psi^{-}\left(  \zeta\right)  }\equiv
G\left(  \zeta\right)  ,\text{ }\zeta\in L_{i},(i=1,2,\cdots m_{L}).
\label{II7.5}%
\end{equation}
According to the homogeneous Riemann-Hilbert problem, from the jump $G\left(
\zeta\right)  $, the fundamental solution $\varphi\left(  \zeta\right)  $ can
be determined:
\begin{equation}
\varphi\left(  \zeta\right)  =e^{\gamma\left(  \zeta\right)  }%
%TCIMACRO{\dprod \limits_{k=1}^{n}}%
%BeginExpansion
{\displaystyle\prod\limits_{k=1}^{n}}
%EndExpansion
\left(  \zeta-c_{k}\right)  ^{\lambda_{k}}, \label{II8}%
\end{equation}
where
\begin{equation}
\gamma\left(  \zeta\right)  =\frac{1}{2\pi i}\int\limits_{L_{1}+L_{2}%
+\cdots+L_{m_{L}}}d\xi\frac{\ln G\left(  \xi\right)  }{\xi-\zeta},
\end{equation}
and the integral is along the boundary, consisting of the arcs $L_{1}$,
$L_{2}$, $\cdots$, and $L_{m_{L}}$. The parameter $\lambda_{k}$ in Eq.
(\ref{II8}) is an integer determined by the following conditions:%
\begin{align}
\mp\operatorname{Re}\frac{\ln G\left(  c_{k}\right)  }{2\pi i}+\lambda_{k}  &
=0,\text{\thinspace\thinspace\thinspace\thinspace\thinspace if\ }%
\mp\operatorname{Re}\frac{\ln G\left(  c_{k}\right)  }{2\pi i}\text{\thinspace
\thinspace is an integer,}\label{II9}\\
-1  &  <\mp\operatorname{Re}\frac{\ln G\left(  c_{k}\right)  }{2\pi i}%
+\lambda_{k}<1,\text{\thinspace\thinspace\thinspace other cases,}
\label{II10a}%
\end{align}
where the upper sign "$-$" has to be taken for the starting point of a certain
arc, $L_{i}$, the lower "$+$" for the end point. The condition (\ref{II10a})
becomes%
\begin{equation}
-1<\mp\operatorname{Re}\frac{\ln G\left(  c_{k}\right)  }{2\pi i}+\lambda
_{k}<0 \label{II10}%
\end{equation}
when the fundamental solution is chosen to satisfy Eq. (\ref{ch}).

By Eq. (\ref{II7.5}), we can see that the jump $G\left(  \zeta\right)  $ of
the fundamental solution $\varphi\left(  \zeta\right)  $ at the two sides of
the boundary is equal to the jump of $\Psi\left(  \zeta\right)  $, a known
function defined by Eqs. (\ref{II1}) and (\ref{II2}), along the boundary. This
allows us to determine the jump $G\left(  \zeta\right)  $ by $\Psi\left(
\zeta\right)  $ and then to determine the fundamental solution $\varphi\left(
\zeta\right)  $ by $G\left(  \zeta\right)  $.

\subsection{Zeros \label{zero}}

The fugacity $z$ is one of the zeros of the function $\Psi\left(
\zeta\right)  $ on the real axis. In principle, after obtaining $\Psi\left(
\zeta\right)  $, except the zero $z$, if we know all other $n_{\omega}-1$
zeros $\omega_{i}$ of $\Psi\left(  \zeta\right)  $, we can write down the
explicit expression for $z$ as in Eq. (\ref{II6}). However, the difficulty of
finding the zeros $\omega_{i}$ is often the same as the difficulty of finding
the zero $z$. That is to say, it is actually impossible to solve $z$ by first
solving the $n_{\omega}-1$ zeros $\omega_{i}$.

Alternatively, to find the zero $z$, we note that Eq. (\ref{II5}) is
essentially an equation of $z$, $\omega_{i}$, and a parameter $\eta$. Based on
Eq. (\ref{II5}), we have two possible ways to construct a set of equations for
$z$, $\omega_{i}$, and $\eta$:

(1) Different values of $\zeta$ give different equations of $z$, $\omega_{i}$,
and $\eta$, and $n_{\omega}+1$ different $\zeta$'s give a set of $n_{\omega
}+1$ equations. Then, solving such a set of equations gives the fugacity $z$.

(2) Deriving both sides of Eq. (\ref{II5}) $\nu$ times ($\nu=0,1,2,\ldots
,n_{\omega}$) and using Eq. (\ref{II7}) give $n_{\omega}+1$ equations of $z $,
$\omega_{i}$, and $\eta$,%
\begin{align}
\left.
%TCIMACRO{\dprod \limits_{k=1}^{n}}%
%BeginExpansion
{\displaystyle\prod\limits_{k=1}^{n}}
%EndExpansion
\left(  \zeta-c_{k}\right)  ^{\beta_{k}}\frac{\Psi\left(  \zeta\right)
}{\varphi\left(  \zeta\right)  }\right\vert _{\zeta=\zeta_{0}}  &  =\left.
\eta\left(  \zeta-z\right)
%TCIMACRO{\dprod \limits_{i=1}^{n_{\omega}-1}}%
%BeginExpansion
{\displaystyle\prod\limits_{i=1}^{n_{\omega}-1}}
%EndExpansion
\left(  \zeta-\omega_{i}\right)  \right\vert _{\zeta=\zeta_{0}},\nonumber\\
\frac{d}{d\zeta}\left[
%TCIMACRO{\dprod \limits_{k=1}^{n}}%
%BeginExpansion
{\displaystyle\prod\limits_{k=1}^{n}}
%EndExpansion
\left(  \zeta-c_{k}\right)  ^{\beta_{k}}\frac{\Psi\left(  \zeta\right)
}{\varphi\left(  \zeta\right)  }\right]  _{\zeta=\zeta_{1}}  &  =\frac
{d}{d\zeta}\left[  \eta\left(  \zeta-z\right)
%TCIMACRO{\dprod \limits_{i=1}^{n_{\omega}-1}}%
%BeginExpansion
{\displaystyle\prod\limits_{i=1}^{n_{\omega}-1}}
%EndExpansion
\left(  \zeta-\omega_{i}\right)  \right]  _{\zeta=\zeta_{1}},\nonumber\\
&  \cdots\cdots\nonumber\\
\frac{d^{n_{\omega}}}{d\zeta^{n_{\omega}}}\left[
%TCIMACRO{\dprod \limits_{k=1}^{n}}%
%BeginExpansion
{\displaystyle\prod\limits_{k=1}^{n}}
%EndExpansion
\left(  \zeta-c_{k}\right)  ^{\beta_{k}}\frac{\Psi\left(  \zeta\right)
}{\varphi\left(  \zeta\right)  }\right]  _{\zeta=\zeta_{n_{\omega}}}  &
=\frac{d^{n_{\omega}}}{d\zeta^{n_{\omega}}}\left[  \eta\left(  \zeta-z\right)
%
%TCIMACRO{\dprod \limits_{i=1}^{n_{\omega}-1}}%
%BeginExpansion
{\displaystyle\prod\limits_{i=1}^{n_{\omega}-1}}
%EndExpansion
\left(  \zeta-\omega_{i}\right)  \right]  _{\zeta=\zeta_{n_{\omega}}}.
\label{equations}%
\end{align}
In our case, it is convenient to chose $\zeta_{0}=\zeta_{1}=\cdots
=\zeta_{n_{\omega}}=0$.

In the following, we will construct the equations for zeros by the second approach.

To find the zeros, we need to construct a set of $n_{\omega}+1$ equations. To
achieve this, we need first to know the value of $n_{\omega}$, the number of
zeros of $\Psi\left(  \zeta\right)  $. In our case, as the function
$\Psi\left(  \zeta\right)  $ has no singularities besides the singularities on
the boundary, the number of the zeros $n_{\omega}$ can be determined with the
help of the argument principle: Along a contour surrounding the complex plane
except the boundary, the change of the argument of $\Psi\left(  \zeta\right)
$ is proportional to the number of zeros.

\section{The explicit expression for the fugacity: isochoric processes
\label{III}}

In this section, we will solve the explicit expression for the fugacity for
weakly interacting Bose and Fermi gases from their equations of state directly
by the method described above. For the weakly interacting case, the
contribution of the collision between gas molecules mainly comes from the
$s$-wave contribution, so the weakly interacting gas can be regarded as a
hard-sphere gas consisting of hard-sphere particles with the scattering length
as its diameter. In the case of Bose gases, we will not consider the problem
of phase transition. It should be noted here that since the equation of state
given by Lee and Yang \cite{LeeYang} is an approximate result for weakly
interacting hard-sphere gases, the expression of the fugacity is, though
explicit, an approximate one limited by the approximation of the equation of state.

\subsection{Three-dimensional hard-sphere Bose gases}

The equation of state for three-dimensional weakly interacting hard-sphere
Bose gases, up to first order of $a/\lambda$, is \cite{LeeYang}%
\begin{align}
\frac{P}{kT}  &  =\frac{1}{V}\ln\Xi\left(  T,z,V\right)  =\left(  2j+1\right)
\frac{1}{\lambda^{3}}\left[  g_{5/2}\left(  z\right)  -2\left(  j+1\right)
\frac{a}{\lambda}g_{3/2}^{2}\left(  z\right)  \right]  ,\label{3DB1}\\
n  &  =\frac{1}{V}z\frac{\partial}{\partial z}\ln\Xi\left(  T,z,V\right)
=\left(  2j+1\right)  \frac{1}{\lambda^{3}}\left[  g_{3/2}\left(  z\right)
-4\left(  j+1\right)  \frac{a}{\lambda}g_{1/2}\left(  z\right)  g_{3/2}\left(
z\right)  \right]  , \label{3DB2}%
\end{align}
where $a$ is the scattering length, $\lambda=h/\sqrt{2\pi mkT}$ the mean
thermal wavelength, $j$ the spin of the particle, and%
\begin{equation}
g_{\sigma}\left(  z\right)  =\frac{1}{\Gamma(\sigma)}\int_{0}^{\infty}%
\frac{t^{\sigma-1}}{z^{-1}e^{t}-1}dt
\end{equation}
the Bose-Einstein integral.

In an isochoric process, the fugacity $z=z\left(  n,T\right)  $ can be solved
from Eq. (\ref{3DB2}). That is to say, as discussed in the above section, the
fugacity $z$ is a zero of the real function $\Psi_{1}\left(  z\right)
=\left(  1/nV\right)  z\left(  \partial/\partial z\right)  \ln\Xi-1$.
Analytically continuing $\Psi_{1}\left(  z\right)  $ to the entire complex
plane gives
\begin{equation}
\Psi_{1}\left(  \zeta\right)  =\left(  2j+1\right)  \frac{1}{n\lambda^{3}%
}\left[  g_{3/2}\left(  \zeta\right)  -4\left(  j+1\right)  \frac{a}{\lambda
}g_{1/2}\left(  \zeta\right)  g_{3/2}\left(  \zeta\right)  \right]  -1,
\label{3DB5}%
\end{equation}
where $g_{\sigma}\left(  \zeta\right)  $ is an analytic continuation of the
Bose-Einstein integral $g_{\sigma}\left(  z\right)  $. Generally speaking,
$\Psi_{1}\left(  \zeta\right)  $ has more than one zero in the complex $\zeta
$-plane, and the fugacity is one of these zeros on the real axis.

By expressing $\Psi_{1}\left(  \zeta\right)  $ as%
\begin{equation}
\Psi_{1}\left(  \zeta\right)  =\eta_{1}\left(  \zeta-z\right)
%TCIMACRO{\dprod \limits_{i=1}^{n_{\omega}-1}}%
%BeginExpansion
{\displaystyle\prod\limits_{i=1}^{n_{\omega}-1}}
%EndExpansion
\left(  \zeta-\omega_{i}\right)  \Phi_{1}\left(  \zeta\right)  , \label{3DB6}%
\end{equation}
where $z$ is the zero corresponding to the fugacity, $\omega_{i}$
($i=1,\cdots,n_{\omega}-1$) are other zeros of $\Psi_{1}\left(  \zeta\right)
$ besides $z$, and $\eta_{1}$ is a constant, we define a function $\Phi
_{1}\left(  \zeta\right)  $ with no zeros. To determine $\Phi_{1}\left(
\zeta\right)  $, we need to analyze the behavior of the singularity of
$\Psi_{1}\left(  \zeta\right)  $ in the $\zeta$-plane. From Eq. (\ref{3DB5}),
we can see that the singularity of $\Psi_{1}\left(  \zeta\right)  $ is
determined by the singularity of the analytically continued Bose-Einstein
integrals $g_{1/2}\left(  \zeta\right)  $ and $g_{3/2}\left(  \zeta\right)  $.
The analytically continued Bose-Einstein integral $g_{\sigma}\left(
\zeta\right)  $ is just the polylogarithm function, or the Jonqui\'{e}re
function, $Li_{\sigma}\left(  \zeta\right)  $, a special case of the Lerch
function, which is analytic in the region with the boundary along the positive
real axis from $1$ to $\infty$ \cite{MOS}. In other words, the analytically
continued Bose-Einstein integral $g_{\sigma}\left(  \zeta\right)  $ has no
isolated singularities, and all the singularities lie on the line from $1$ to
$\infty$ on the real axis, i.e., the boundary is $\left[  1,\infty\right)  $.
Consequently, according to Eqs. (\ref{3DB5}) and (\ref{3DB6}), $\Psi
_{1}\left(  \zeta\right)  $ and $\Phi_{1}\left(  \zeta\right)  $ also have no
isolated singularities, and all their singularities lie on the line $L$ from
$1$ to $\infty$ on the real axis, i.e., in this case, the boundary of the
analytic region consists of only one line. The function $\Phi_{1}\left(
\zeta\right)  $ is analytic in the region with the boundary $L$ and everywhere
different from zero.

Calculating $\Phi_{1}\left(  \zeta\right)  $ needs the solution of the
homogeneous Riemann-Hilbert problem. Eq. (\ref{II7}) gives the relation
between $\Phi_{1}\left(  \zeta\right)  $ and the fundamental solution of the
homogeneous Riemann-Hilbert problem $\varphi_{1}\left(  \zeta\right)  $. As we
have chosen the fundamental solution satisfying Eq. (\ref{ch}), the parameter
$\beta_{k}$ in Eq. (\ref{II7}) can be determined by the demand that the choice
of $\beta_{k}$ must ensure that the behaviors of both sides of Eq. (\ref{II7})
are the same at each end. In our problem, the boundary of the analytic region
of $\Psi_{1}\left(  \zeta\right)  $ has only one end $\zeta=1$ besides the end
at infinity. Near the end $\zeta=1$,%
\begin{equation}
\left.  \Psi_{1}\left(  \zeta\right)  \right\vert _{\zeta\rightarrow1}%
\sim\frac{1}{\sqrt{\zeta-1}},
\end{equation}
i.e., $\Psi_{1}\left(  \zeta\right)  $ is divergent near $\zeta=1$ with a
degree less than $1$. This means $\beta=0$. The singularities of $\Phi
_{1}\left(  \zeta\right)  $ and $\Psi_{1}\left(  \zeta\right)  $ are the same,
so $\Phi_{1}\left(  \zeta\right)  $ itself is just the fundamental solution,
i.e., $\Phi_{1}\left(  \zeta\right)  =\varphi_{1}\left(  \zeta\right)  $.

To solve the fundamental solution $\varphi_{1}\left(  \zeta\right)  $, we
first need to calculate the jump,%
\begin{equation}
G_{1}\left(  x\right)  =\frac{\varphi_{1}^{+}\left(  x\right)  }{\varphi
_{1}^{-}\left(  x\right)  }=\frac{\Psi_{1}^{+}\left(  x\right)  }{\Psi_{1}%
^{-}\left(  x\right)  },\text{ }x\in\left[  1,\infty\right)  ,
\end{equation}
on the boundary, according to Eq. (\ref{II7.5}).

The jump of $\Psi_{1}\left(  \zeta\right)  $ on the boundary, according to Eq.
(\ref{3DB5}), is determined by the jump of the analytically continued
Bose-Einstein integral $g_{\sigma}\left(  \zeta\right)  $, the polylogarithm
function $Li_{\sigma}\left(  \zeta\right)  $. The imaginary part of the
polylogarithm function $Li_{\sigma}\left(  \zeta\right)  $ has a discontinuity
on the boundary $\left[  1,\infty\right)  $ \cite{Wood}:%
\begin{equation}
\operatorname{Im}Li_{\sigma}\left(  x\pm i\delta\right)  =\pm\frac{\pi}%
{\Gamma\left(  \sigma\right)  }\left(  \ln x\right)  ^{\sigma-1},\text{ }%
x\in\left[  1,\infty\right)  , \label{Li}%
\end{equation}
where $\delta$ is a small positive quantity. Therefore, the values of the
analytically continued Bose-Einstein integral at two sides of the boundary are%
\begin{equation}
g_{\sigma}^{\pm}\left(  x\right)  =Li_{\sigma}^{\pm}\left(  x\right)
=\mathfrak{g}_{\sigma}\left(  x\right)  \pm i\frac{\pi}{\Gamma\left(
\sigma\right)  }\left(  \ln x\right)  ^{\sigma-1},\text{ \ }x\in\left[
1,\infty\right)  , \label{BEICP}%
\end{equation}
where%
\begin{equation}
\mathfrak{g}_{\sigma}\left(  x\right)  \equiv\frac{1}{\Gamma(\sigma
)}\mathbb{P}\int_{0}^{\infty}\frac{t^{\sigma-1}}{x^{-1}e^{t}-1}dt,\text{
}\sigma\neq0
\end{equation}
denotes the Cauchy principal value of the analytically continued Bose-Einstein
integral $g_{\sigma}\left(  \zeta\right)  $ at the point $x$ on the boundary
\cite{Clunie}. Note that the Bose-Einstein integral $g_{0}\left(  x\right)  $,
i.e., the case of $\sigma=0$, is an exception. $g_{0}\left(  \zeta\right)
=\zeta/\left(  1-\zeta\right)  $ has only one singularity $\zeta=1$, but has
no singularities on the region $\left(  1,\infty\right)  $ on the real axis.
Accordingly, by Eq. (\ref{3DB5}), the values of $\Psi_{1}\left(  \zeta\right)
$ on both sides of the boundary $\left[  1,\infty\right)  $ are
\begin{align}
\Psi_{1}^{\pm}\left(  x\right)   &  =\left(  2j+1\right)  \frac{1}%
{n\lambda^{3}}\left\{  \mathfrak{g}_{3/2}\left(  x\right)  -4\left(
j+1\right)  \frac{a}{\lambda}\left[  \mathfrak{g}_{1/2}\left(  x\right)
\mathfrak{g}_{3/2}\left(  x\right)  -2\pi\right]  \right\}  -1\nonumber\\
&  \pm i\left(  2j+1\right)  \frac{1}{n\lambda^{3}}2\sqrt{\pi}\sqrt{\ln
x}\left\{  1-2\left(  j+1\right)  \frac{a}{\lambda}\left[  \frac{1}{\ln
x}\mathfrak{g}_{3/2}\left(  x\right)  +2\mathfrak{g}_{1/2}\left(  x\right)
\right]  \right\}  .
\end{align}
Clearly, $\Psi_{1}^{+}\left(  x\right)  $ and $\Psi_{1}^{-}\left(  x\right)  $
are complex conjugate to each other, i.e., $\Psi_{1}^{+}\left(  x\right)
=\left[  \Psi_{1}^{-}\left(  x\right)  \right]  ^{\ast}$, and the jump on the
boundary $\left[  1,\infty\right)  $ is then%
\begin{equation}
G_{1}\left(  x\right)  =\exp\left[  i2\arg\Psi_{1}^{+}\left(  x\right)
\right]  ,
\end{equation}
where the argument of $\Psi_{1}^{+}\left(  x\right)  $ is%
\begin{equation}
\arg\Psi_{1}^{+}\left(  x\right)  =\operatorname{arccot}\frac
{\operatorname{Re}\Psi_{1}^{+}\left(  x\right)  }{\operatorname{Im}\Psi
_{1}^{+}\left(  x\right)  }. \label{arg1}%
\end{equation}

Now, the fundamental solution $\varphi_{1}\left(  \zeta\right)  $ can be
solved by use of Eq. (\ref{II8}) directly. In our problem, the boundary has
only one end different from infinity, $\zeta=1$, which means%
\begin{equation}
\varphi_{1}\left(  \zeta\right)  =e^{\gamma_{1}\left(  \zeta\right)  }\left(
\zeta-1\right)  ^{\lambda_{1}},
\end{equation}
where
\begin{equation}
\gamma_{1}\left(  \zeta\right)  =\frac{1}{\pi}%
%TCIMACRO{\dint \nolimits_{1}^{\infty}}%
%BeginExpansion
{\displaystyle\int\nolimits_{1}^{\infty}}
%EndExpansion
dx\frac{\arg\Psi_{1}^{+}\left(  x\right)  }{x-\zeta}%
\end{equation}
and $\lambda_{1}$ is determined by Eq. (\ref{II10}). Choosing $\arg\Psi
_{1}^{+}\left(  \infty\right)  =0$, we have $\arg\Psi_{1}^{+}\left(  1\right)
=-3\pi/2$. As the fundamental solution satisfies Eq. (\ref{ch}), the relation
Eq. (\ref{II10}) becomes%
\begin{equation}
-1<-\frac{1}{\pi}\arg\Psi_{1}^{+}\left(  1\right)  +\lambda_{1}<0,
\end{equation}
and gives $\lambda_{1}=-2$. Therefore,
\begin{equation}
\Phi_{1}\left(  \zeta\right)  =\varphi_{1}\left(  \zeta\right)  =\frac
{e^{\gamma_{1}\left(  \zeta\right)  }}{\left(  \zeta-1\right)  ^{2}}.
\label{3DPHI1}%
\end{equation}

By Eq. (\ref{3DB6}), we can construct a set of equations for zeros. For this
purpose, we need to determine the number of the zeros of $\Psi_{1}\left(
\zeta\right)  $. The fact that the function $\Psi_{1}\left(  \zeta\right)  $
has no singularities besides the boundary $\left[  1,\infty\right)  $ allows
us to use the argument principle to determine the number of its zeros in the
complex $\zeta$-plane directly. Applying the argument principle along the
contour surrounding the complex plane except the boundary $\left[
1,\infty\right)  $ shows that $\Psi_{1}\left(  \zeta\right)  $ has two zeros,
denoted as $z$ (the fugacity) and $\omega$. Substituting Eq. (\ref{3DPHI1})
into Eq. (\ref{3DB6})\ with $n_{\omega}=2$ gives%
\begin{equation}
\Psi_{1}\left(  \zeta\right)  =\eta_{1}\frac{\left(  \zeta-z\right)  \left(
\zeta-\omega\right)  }{\left(  \zeta-1\right)  ^{2}}e^{\gamma_{1}\left(
\zeta\right)  }. \label{zero1}%
\end{equation}
From Eq. (\ref{zero1}), we can construct a set of equations for $z$, $\omega$,
and $\eta_{1}$. Since in this case, the values of the functions in Eq.
(\ref{zero1}) are relatively easy to be carried out at $\zeta=0$, we adopt the
second method introduced in Sec. \ref{zero}: Derive both sides of Eq.
(\ref{zero1}) to construct various equations for zeros. In this case, we need
three equations for determining $z$, $\omega$, and $\eta_{1}$. When $\zeta=0$,
by Eq. (\ref{equations}), we have%
\begin{equation}
\left\{
\begin{array}
[c]{l}%
%TCIMACRO{\TeXButton{displaystyle}{\displaystyle}}%
%BeginExpansion
\displaystyle
%EndExpansion
\eta_{1}z\omega e^{\gamma_{1}\left(  0\right)  }=\Psi_{1}\left(  0\right)  ,\\%
%TCIMACRO{\TeXButton{displaystyle}{\displaystyle}}%
%BeginExpansion
\displaystyle
%EndExpansion
\eta_{1}z\omega e^{\gamma_{1}\left(  0\right)  }\left[  \gamma_{1}^{\prime
}\left(  0\right)  -\frac{1}{z}-\frac{1}{\omega}+2\right]  =\Psi_{1}^{\prime
}\left(  0\right)  ,\\%
%TCIMACRO{\TeXButton{displaystyle}{\displaystyle}}%
%BeginExpansion
\displaystyle
%EndExpansion
\eta_{1}z\omega e^{\gamma_{1}\left(  0\right)  }\left\{  \gamma_{1}^{\prime
}\left(  0\right)  ^{2}-2\left[  \gamma_{1}^{\prime}\left(  0\right)
+2\right]  \left(  \frac{1}{z}+\frac{1}{\omega}\right)  +4\gamma_{1}^{\prime
}\left(  0\right)  +\gamma_{1}^{\prime\prime}\left(  0\right)  +\frac
{2}{z\omega}+6\right\}  =\Psi_{1}^{\prime\prime}\left(  0\right)  ,
\end{array}
\right.  \label{3Dzero}%
\end{equation}
where%
\begin{equation}
\gamma_{1}^{\left(  n\right)  }\left(  0\right)  =\frac{n!}{\pi}\int
_{1}^{\infty}\frac{\arg\Psi_{1}^{+}\left(  x\right)  }{x^{n+1}}dx.
\end{equation}
Consequently, the fugacity $z$ can be obtained by solving Eq. (\ref{3Dzero}):%
\begin{align}
z  &  =2\frac{n\lambda^{3}}{2j+1}\left\{  1+\frac{n\lambda^{3}}{2j+1}\left[
\gamma_{1}^{\prime}\left(  0\right)  +2\right]  +\left\{  1-2\frac
{n\lambda^{3}}{2j+1}\left[  \gamma_{1}^{\prime}\left(  0\right)
+2-\frac{\sqrt{2}}{2}+8\left(  j+1\right)  \frac{a}{\lambda}\right]  \right.
\right. \nonumber\\
&  \left.  \left.  -\frac{\left(  n\lambda^{3}\right)  ^{2}}{\left(
2j+1\right)  ^{2}}\left[  4\gamma_{1}^{\prime}\left(  0\right)  +\gamma
_{1}^{\prime}\left(  0\right)  ^{2}-2\gamma_{1}^{\prime\prime}\left(
0\right)  \right]  \right\}  ^{1/2}\right\}  ^{-1}. \label{3DBEVz}%
\end{align}

Substituting the fugacity $z$ given by Eq. (\ref{3DBEVz}) into Eq.
(\ref{3DB1}) gives the equation of state. Note that the fugacity here is an
explicit function of the temperature $T$.

To illustrate the above result more clearly, we plot the fugacity as a
function of temperature in figure 1. Since eq. (\ref{3DBEVz}) is an exact
solution of eq. (\ref{3DB2}), the result is exactly the same as the numerical
solution of eq. (\ref{3DB2}). Note that at this situation, the fugacity has a
maximum value, i.e., the curve in figure 1 has an end.

\begin{figure}[ptb]
\begin{center}
\includegraphics[height=2.5in]
{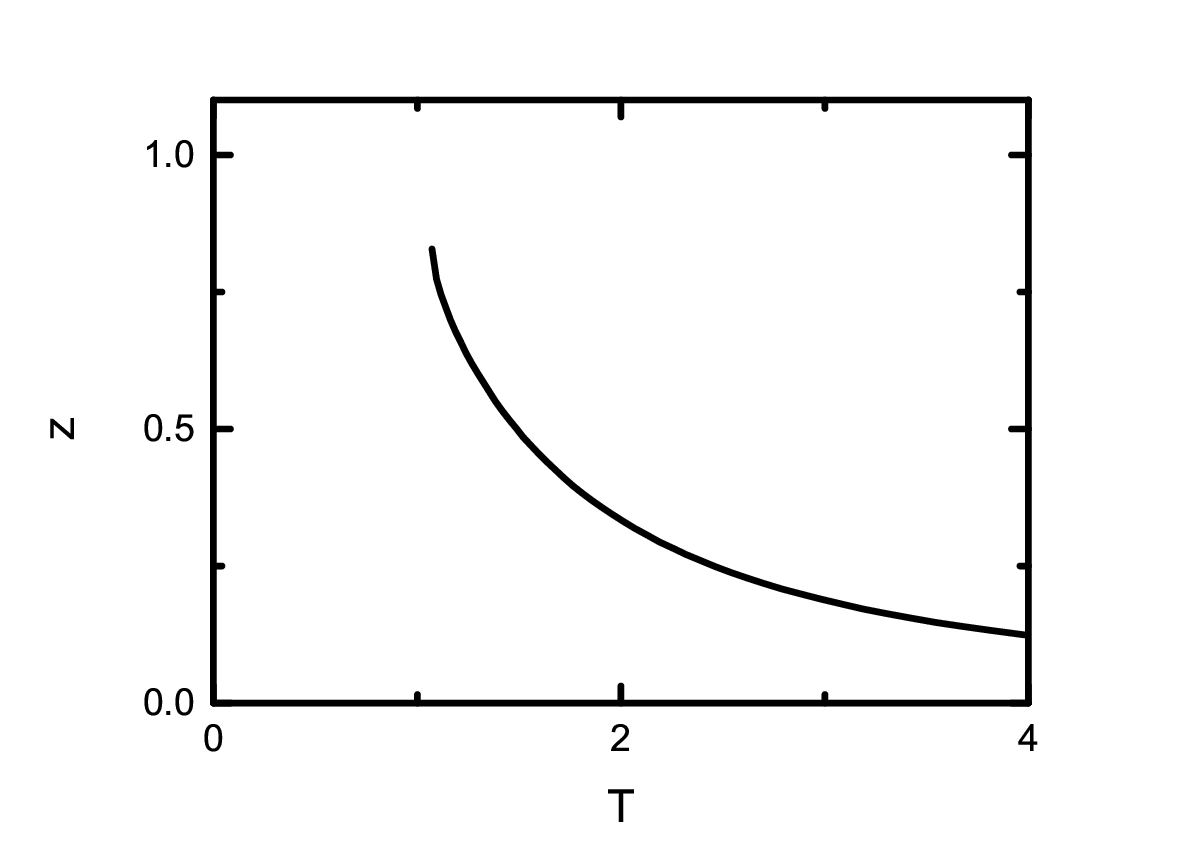}
\end{center}
\caption{The relation between $z$ and $T$ for three-dimensional hard-sphere
Bose gases in an isochoric process.}%
\label{fig1}%
\end{figure}

\subsection{Three-dimensional hard-sphere Fermi gases}

The equation of state for three-dimensional weakly interacting hard-sphere
Fermi gases, up to first order of $a/\lambda$, is \cite{LeeYang}
\begin{align}
\frac{P}{kT}  &  =\left(  2j+1\right)  \frac{1}{\lambda^{3}}\left[
f_{5/2}\left(  z\right)  -2j\frac{a}{\lambda}f_{3/2}^{2}\left(  z\right)
\right]  ,\label{3DFDP}\\
n  &  =\left(  2j+1\right)  \frac{1}{\lambda^{3}}\left[  f_{3/2}\left(
z\right)  -4j\frac{a}{\lambda}f_{1/2}\left(  z\right)  f_{3/2}\left(
z\right)  \right]  , \label{3DFDn}%
\end{align}
where the Fermi-Dirac integral%
\begin{equation}
f_{\sigma}\left(  z\right)  =\frac{1}{\Gamma(\sigma)}\int_{0}^{\infty}%
\frac{t^{\sigma-1}}{z^{-1}e^{t}+1}dt.
\end{equation}

In an isochoric process, the fugacity $z=z\left(  n,T\right)  $ can be solved
from Eq. (\ref{3DFDn}). The fugacity $z$ is a zero of the complex function%

\begin{equation}
\Psi_{2}\left(  \zeta\right)  =\left(  2j+1\right)  \frac{1}{n\lambda^{3}%
}\left[  f_{3/2}\left(  \zeta\right)  -4j\frac{a}{\lambda}f_{1/2}\left(
\zeta\right)  f_{3/2}\left(  \zeta\right)  \right]  -1 \label{5.010}%
\end{equation}
on the real axis, where $f_{\sigma}\left(  \zeta\right)  $ is the analytically
continued Fermi-Dirac integral. The singularity of $\Psi_{2}\left(
\zeta\right)  $ is determined by the singularity of the analytically continued
Fermi-Dirac integral. The analytically continued Fermi-Dirac integral here is
the polylogarithm function $-Li_{\sigma}\left(  -\zeta\right)  $, which is
analytic in the region with the boundary $\left[  -1,-\infty\right)  $.

By the argument principle, along the contour surrounding the complex plane
except the boundary $\left[  -1,-\infty\right)  $, we can determine that
$\Psi_{2}\left(  \zeta\right)  $ has two real zeros on the real axis; one of
these two zeros is just the fugacity $z$. Accordingly, $\Psi_{2}\left(
\zeta\right)  $ can be expressed as
\begin{equation}
\Psi_{2}\left(  \zeta\right)  =\eta_{2}\left(  \zeta-z\right)  \left(
\zeta-\omega\right)  \Phi_{2}\left(  \zeta\right)  , \label{3DF6}%
\end{equation}
where the function $\Phi_{2}\left(  \zeta\right)  $ has the same singularities
as those of $\Psi_{2}\left(  \zeta\right)  $, lying on the line $\left[
-1,-\infty\right)  $, and has no zeros.

As stated above, in our case, the fundamental solution is chosen to satisfy
Eq. (\ref{ch}). Near the end different from infinity of the boundary $\left[
-1,-\infty\right)  $, $\zeta=-1$, we have%
\begin{equation}
\left.  \Psi_{2}\left(  \zeta\right)  \right\vert _{\zeta\rightarrow-1}%
\sim\frac{1}{\sqrt{\zeta+1}}.
\end{equation}
The degree of divergence of $\Psi_{2}\left(  \zeta\right)  $ at the end
$\zeta=-1$ being less than $1$ implies that the function $\Phi_{2}\left(
\zeta\right)  $ itself is a fundamental solution of the homogeneous
Riemann-Hilbert problem, i.e., $\Phi_{2}\left(  \zeta\right)  =\varphi
_{2}\left(  \zeta\right)  $, where $\varphi_{2}\left(  \zeta\right)  $ denotes
the fundamental solution.

To solve the fundamental solution $\varphi_{2}\left(  \zeta\right)  $, we need
to know the jump on the boundary. From Eq. (\ref{II7.5}), we have
\begin{equation}
G_{2}\left(  x\right)  =\frac{\varphi_{2}^{+}\left(  x\right)  }{\varphi
_{2}^{-}\left(  x\right)  }=\frac{\Psi_{2}^{+}\left(  x\right)  }{\Psi_{2}%
^{-}\left(  x\right)  },\text{ }x\in\left[  -1,-\infty\right)  ,
\end{equation}
and from Eq. (\ref{Li}), we have%

\begin{equation}
f_{\sigma}^{\pm}\left(  x\right)  =-Li_{\sigma}^{\pm}\left(  -x\right)
=\mathfrak{f}_{\sigma}\left(  x\right)  \mp i\frac{\pi}{\Gamma\left(
\sigma\right)  }\left[  \ln\left(  -x\right)  \right]  ^{\sigma-1},
\label{FDI}%
\end{equation}
where%
\begin{equation}
\mathfrak{f}_{\sigma}\left(  x\right)  =\frac{1}{\Gamma(\sigma)}\mathbb{P}%
\int_{0}^{\infty}\frac{t^{\sigma-1}}{x^{-1}e^{t}+1}dt,\text{ }\sigma\neq0
\label{PFDI}%
\end{equation}
is the Cauchy principal value of the analytically continued Fermi-Dirac
integral. Notice that $x<0$. Then, the value of $\Psi_{2}\left(  \zeta\right)
$ at the two sides of the boundary can be obtained directly:%
\begin{align}
\Psi_{2}^{\pm}\left(  x\right)   &  =\left(  2j+1\right)  \frac{1}%
{n\lambda^{3}}\left\{  \mathfrak{f}_{3/2}\left(  x\right)  -4j\frac{a}%
{\lambda}\left[  \mathfrak{f}_{1/2}\left(  x\right)  \mathfrak{f}_{3/2}\left(
x\right)  -2\pi\right]  \right\}  -1\nonumber\\
&  \mp i\left(  2j+1\right)  \frac{1}{n\lambda^{3}}2\sqrt{\pi}\sqrt{\ln\left(
-x\right)  }\left\{  1-2j\frac{a}{\lambda}\left[  \frac{1}{\ln\left(
-x\right)  }\mathfrak{f}_{3/2}\left(  x\right)  +2\mathfrak{f}_{1/2}\left(
x\right)  \right]  \right\}  . \label{psi2}%
\end{align}
$\Psi_{2}^{+}\left(  x\right)  $ and $\Psi_{2}^{-}\left(  x\right)  $ are
complex conjugate to each other, so the jump of $\Psi_{2}\left(  x\right)  $
on the boundary $\left[  -1,-\infty\right)  $ is
\begin{equation}
G_{2}\left(  x\right)  =\exp\left[  i2\arg\Psi_{2}^{+}\left(  x\right)
\right]  ,
\end{equation}
where the argument $\arg\Psi_{2}^{+}\left(  x\right)  =\operatorname{arccot}%
\left[  \operatorname{Re}\Psi_{2}^{+}\left(  x\right)  /\operatorname{Im}%
\Psi_{2}^{+}\left(  x\right)  \right]  $. Noting that in this case, there is
only one end different from infinity of the boundary, $\zeta=-1$, we can write
down the fundamental solution from Eq. (\ref{II8}):%
\begin{equation}
\varphi_{2}\left(  \zeta\right)  =e^{\gamma_{2}\left(  \zeta\right)  }\left(
\zeta+1\right)  ^{\lambda_{2}},
\end{equation}
where%
\begin{equation}
\gamma_{2}\left(  \zeta\right)  =\frac{1}{\pi}%
%TCIMACRO{\dint \nolimits_{-1}^{-\infty}}%
%BeginExpansion
{\displaystyle\int\nolimits_{-1}^{-\infty}}
%EndExpansion
dx\frac{\arg\Psi_{2}^{+}\left(  x\right)  }{x-\zeta} \label{gamma2}%
\end{equation}
and $\lambda_{2}$ will be determined by the condition (\ref{II10}). Choosing
$\arg\Psi_{2}^{+}\left(  -\infty\right)  =0$ gives $\arg\Psi_{2}^{+}\left(
-1\right)  =-3\pi/2$. As the fundamental solution satisfies Eq. (\ref{ch}),
the condition (\ref{II10}) becomes%
\begin{equation}
-1<-\frac{1}{\pi}\arg\Psi_{2}^{+}\left(  -1\right)  +\lambda_{2}<0,
\end{equation}
and gives $\lambda_{2}=-2$. Then,%

\begin{equation}
\Phi_{2}\left(  \zeta\right)  =\varphi_{2}\left(  \zeta\right)  =\frac
{e^{\gamma_{2}\left(  \zeta\right)  }}{\left(  \zeta+1\right)  ^{2}}.
\label{Phi2}%
\end{equation}
Substituting Eq. (\ref{Phi2}) into Eq. (\ref{3DF6}), we have%
\begin{equation}
\Psi_{2}\left(  \zeta\right)  =\eta_{2}\frac{\left(  \zeta-z\right)  \left(
\zeta-\omega\right)  }{\left(  \zeta+1\right)  ^{2}}e^{\gamma_{2}\left(
\zeta\right)  }. \label{zero2}%
\end{equation}
To determine the zeros $z$, $\omega$ and the constant $\eta_{2}$, we need
three equations. By setting $\zeta=0$, Eq. (\ref{zero2}) and its first- and
second-order derivatives give these three equations. Consequently, we have%
\begin{align}
z  &  =2\frac{n\lambda^{3}}{2j+1}\left\{  1+\frac{n\lambda^{3}}{2j+1}\left[
\gamma_{2}^{\prime}\left(  0\right)  -2\right]  +\left\{  1-2\frac
{n\lambda^{3}}{2j+1}\left[  \gamma_{2}^{\prime}\left(  0\right)
-2+\frac{\sqrt{2}}{2}+8j\frac{a}{\lambda}\right]  \right.  \right. \nonumber\\
&  \left.  \left.  +\frac{\left(  n\lambda^{3}\right)  ^{2}}{\left(
2j+1\right)  ^{2}}\left[  4\gamma_{2}^{\prime}\left(  0\right)  -\gamma
_{2}^{\prime}\left(  0\right)  ^{2}+2\gamma_{2}^{\prime\prime}\left(
0\right)  \right]  \right\}  ^{1/2}\right\}  ^{-1}. \label{3DFDVz}%
\end{align}
This relation between $z$\ and $T$\ is plotted in figure 2.

\begin{figure}[ptb]
\begin{center}
\includegraphics[height=2.5in]
{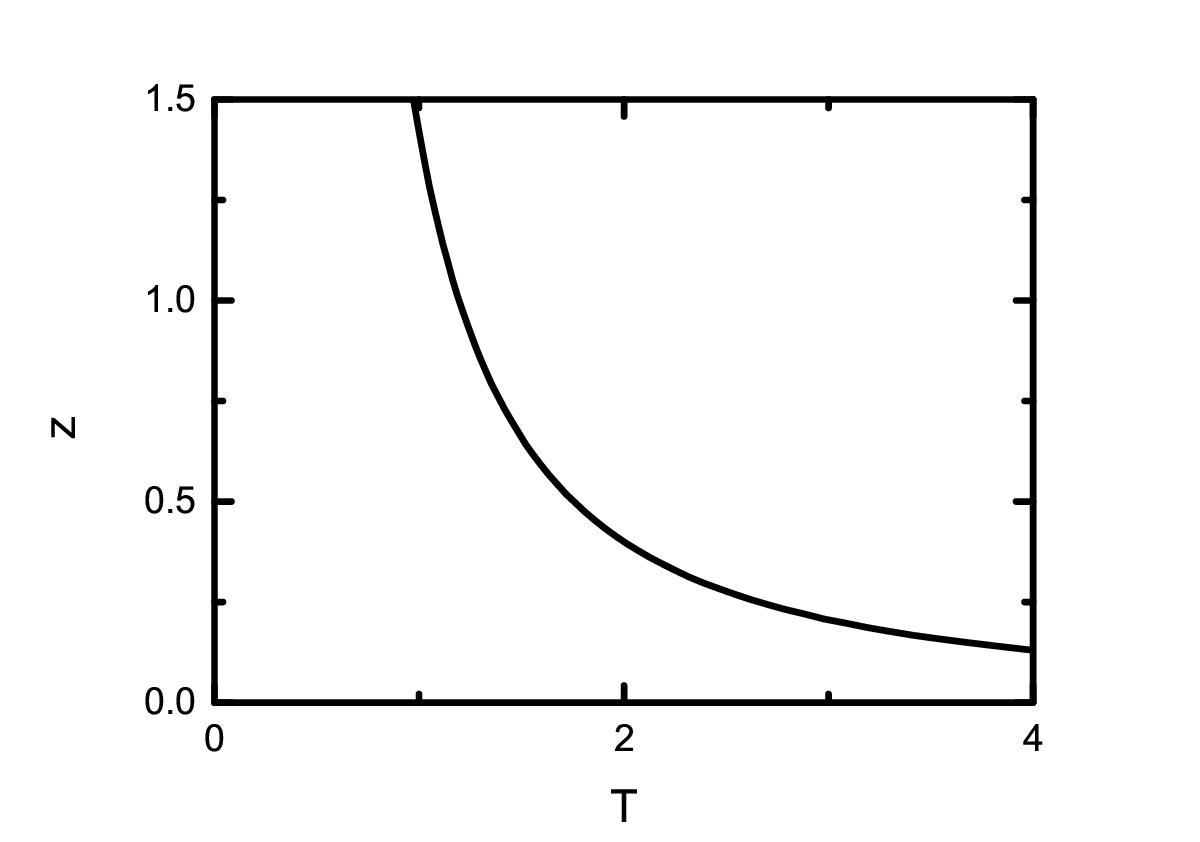}
\end{center}
\caption{The relation between $z$ and $T$ for three-dimensional hard-sphere
Fermi gases in an isochoric process.}%
\label{fig2}%
\end{figure}

Substituting the fugacity $z$ given by Eq. (\ref{3DFDVz}) into Eq.
(\ref{3DFDP}) gives the equation of state.

\subsection{Two-dimensional hard-sphere Bose gases}

The equation of state for two-dimensional weakly interacting hard-sphere Bose
gases, up to first order of $a/\lambda$, is \cite{OurEPL}%
\begin{align}
\frac{P}{kT}  &  =\left(  2j+1\right)  \frac{1}{\lambda^{2}}\left[
g_{2}\left(  z\right)  -\left(  j+1\right)  \pi\frac{a}{\lambda}g_{1}\left(
z\right)  g_{3/2}\left(  z\right)  \right]  ,\label{2DBEP}\\
n  &  =\left(  2j+1\right)  \frac{1}{\lambda^{2}}\left\{  g_{1}\left(
z\right)  -\left(  j+1\right)  \pi\frac{a}{\lambda}\left[  g_{0}\left(
z\right)  g_{3/2}\left(  z\right)  +g_{1/2}\left(  z\right)  g_{1}\left(
z\right)  \right]  \right\}  , \label{2DBEn}%
\end{align}
where $n$ is the area density. In an isochoric process, the fugacity $z$ can
be solved from Eq. (\ref{2DBEn}). Following the above analysis, $z$ is a zero
of the complex function
\begin{equation}
\Psi_{3}\left(  \zeta\right)  =\left(  2j+1\right)  \frac{1}{n\lambda^{2}%
}\left\{  g_{1}\left(  \zeta\right)  -\left(  j+1\right)  \pi\frac{a}{\lambda
}\left[  g_{0}\left(  \zeta\right)  g_{3/2}\left(  \zeta\right)
+g_{1/2}\left(  \zeta\right)  g_{1}\left(  \zeta\right)  \right]  \right\}  -1
\label{2DB5}%
\end{equation}
on the real axis. Note that $g_{\sigma}\left(  \zeta\right)  $ is an analytic
continuation of the Bose-Einstein integral.

The singularities of $\Psi_{3}\left(  \zeta\right)  $ are determined by the
singularities of the analytically continued Bose-Einstein integral $g_{\sigma
}\left(  \zeta\right)  $, the polylogarithm function $Li_{\sigma}\left(
\zeta\right)  $. $\Psi_{3}\left(  \zeta\right)  $ has no isolated
singularities, and all of its singularities form a line from $1$ to $\infty$
on the real axis. Similar to the three-dimensional case, by use of the
argument principle, we can find that $\Psi_{3}\left(  \zeta\right)  $ has two
zeros in the complex $\zeta$-plane, denoted as $z$ and $\omega$, where $z $ is
the fugacity. As a result, $\Psi_{3}\left(  \zeta\right)  $ can be expressed
as%
\begin{equation}
\Psi_{3}\left(  \zeta\right)  =\eta_{3}\left(  \zeta-z\right)  \left(
\zeta-\omega\right)  \Phi_{3}\left(  \zeta\right)  , \label{2DB6}%
\end{equation}
where the function $\Phi_{3}\left(  \zeta\right)  $ has no zeros, and its
singularities also lie on the line $\left[  1,\infty\right)  $.

Near the end point $\zeta=1$ of the boundary of the analytic region of
$\Psi_{3}\left(  \zeta\right)  $,
\begin{equation}
\left.  \Psi_{3}\left(  \zeta\right)  \right\vert _{\zeta\rightarrow1}%
\sim\frac{1}{\zeta-1},
\end{equation}
i.e., $\Psi_{3}\left(  \zeta\right)  $ is linearly divergent near the point of
$1$. As the singularities of $\Phi_{3}\left(  \zeta\right)  $ and $\Psi
_{3}\left(  \zeta\right)  $ are the same, $\Phi_{3}\left(  \zeta\right)  $ is
not a fundamental solution of the homogeneous Riemann-Hilbert problem.
According to Eq. (\ref{II7}), we choose $\beta=1$ and express $\Phi_{3}\left(
\zeta\right)  $ as%
\begin{equation}
\Phi_{3}\left(  \zeta\right)  =\frac{\varphi_{3}\left(  \zeta\right)  }%
{\zeta-1}, \label{2DB8}%
\end{equation}
where $\varphi_{3}\left(  \zeta\right)  $ is a fundamental solution. To find
the fundamental solution $\varphi_{3}\left(  \zeta\right)  $, we need the jump
of $\Psi_{3}\left(  \zeta\right)  $ on the boundary $\left[  1,\infty\right)
$,%
\begin{equation}
G_{3}\left(  x\right)  =\frac{\varphi_{3}^{+}\left(  x\right)  }{\varphi
_{3}^{-}\left(  x\right)  }=\frac{\Psi_{3}^{+}\left(  x\right)  }{\Psi_{3}%
^{-}\left(  x\right)  },\text{ }x\in\left[  1,\infty\right)  .
\end{equation}
Similar to the three-dimensional case, the values of $\Psi_{3}\left(
\zeta\right)  $ at the two sides of the boundary $\left[  1,\infty\right)  $
are easy to be obtained from Eq. (\ref{2DB5}) by use of Eq. (\ref{BEICP}):%
\begin{align}
\Psi_{3}^{\pm}\left(  x\right)   &  =\left(  2j+1\right)  \frac{1}%
{n\lambda^{2}}\left\{  \mathfrak{g}_{1}\left(  x\right)  -\left(  j+1\right)
\pi\frac{a}{\lambda}\left[  g_{0}\left(  x\right)  \mathfrak{g}_{3/2}\left(
x\right)  +\mathfrak{g}_{1/2}\left(  x\right)  \mathfrak{g}_{1}\left(
x\right)  -\frac{\pi^{3/2}}{\sqrt{\ln x}}\right]  \right\}  -1\nonumber\\
&  \pm i\left(  2j+1\right)  \frac{1}{n\lambda^{2}}\pi\left\{  1-\left(
j+1\right)  \sqrt{\pi}\frac{a}{\lambda}\left[  2g_{0}\left(  x\right)
\sqrt{\ln x}+\frac{1}{\sqrt{\ln x}}\mathfrak{g}_{1}\left(  x\right)
+\sqrt{\pi}\mathfrak{g}_{1/2}\left(  x\right)  \right]  \right\}  .
\end{align}
Since $\left[  \Psi_{3}^{+}\left(  x\right)  \right]  ^{\ast}=\Psi_{3}%
^{-}\left(  x\right)  $, the jump on the boundary $\left[  1,\infty\right)  $
is%
\begin{equation}
G_{3}\left(  x\right)  =\exp\left[  i2\arg\Psi_{3}^{+}\left(  x\right)
\right]  ,
\end{equation}
where the argument $\arg\Psi_{3}^{+}\left(  x\right)  =\operatorname{arccot}%
\left[  \operatorname{Re}\Psi_{3}^{+}\left(  x\right)  /\operatorname{Im}%
\Psi_{3}^{+}\left(  x\right)  \right]  $.

The fundamental solution can be obtained from Eq. (\ref{II8}). In our problem,
$\zeta=1$ is the unique end of the boundary different from infinity, so we
have%
\begin{equation}
\varphi_{3}\left(  \zeta\right)  =e^{\gamma_{3}\left(  \zeta\right)  }\left(
\zeta-1\right)  ^{\lambda_{3}},
\end{equation}
where%
\begin{equation}
\gamma_{3}\left(  \zeta\right)  =\frac{1}{\pi}%
%TCIMACRO{\dint \nolimits_{1}^{\infty}}%
%BeginExpansion
{\displaystyle\int\nolimits_{1}^{\infty}}
%EndExpansion
dx\frac{\arg\Psi_{3}^{+}\left(  x\right)  }{x-\zeta}.
\end{equation}
In our problem, the fundamental solution is chosen to satisfy the condition
(\ref{ch}). Choosing $\arg\Psi_{3}^{+}\left(  \infty\right)  =0$ gives
$\arg\Psi_{3}^{+}\left(  1\right)  =-\pi$. Then, Eq. (\ref{II9}) becomes%
\begin{equation}
-\frac{1}{\pi}\arg\Psi_{3}^{+}\left(  x\right)  +\lambda_{3}=0,
\end{equation}
and gives $\lambda_{3}=-1$, i.e.,%

\begin{equation}
\varphi_{3}\left(  \zeta\right)  =\frac{e^{\gamma_{3}\left(  \zeta\right)  }%
}{\zeta-1}.
\end{equation}
From Eqs. (\ref{2DB6}) and (\ref{2DB8}), we have%
\begin{equation}
\Psi_{3}\left(  \zeta\right)  =\eta_{3}\frac{\left(  \zeta-z\right)  \left(
\zeta-\omega\right)  }{\left(  \zeta-1\right)  ^{2}}e^{\gamma_{3}\left(
\zeta\right)  }. \label{zero3}%
\end{equation}
By setting $\zeta=0$, Eq. (\ref{zero3}) and its first- and the second-order
derivatives give three equations for $z$, $\omega$, and the constant $\eta
_{3}$. Solving this set of equations gives the fugacity%
\begin{align}
z  &  =2\frac{n\lambda^{2}}{\left(  2j+1\right)  }\left\{  1+\frac
{n\lambda^{2}}{\left(  2j+1\right)  }\left[  \gamma_{3}^{\prime}\left(
0\right)  +2\right]  +\left\{  1-2\frac{n\lambda^{2}}{\left(  2j+1\right)
}\left[  \gamma_{3}^{\prime}\left(  0\right)  +1+4\pi\left(  j+1\right)
\frac{a}{\lambda}\right]  \right.  \right. \nonumber\\
&  \left.  \left.  -\frac{\left(  n\lambda^{2}\right)  ^{2}}{\left(
2j+1\right)  }\left[  4\gamma_{3}^{\prime}\left(  0\right)  +\gamma
_{3}^{\prime}\left(  0\right)  ^{2}-2\gamma_{3}^{\prime\prime}\left(
0\right)  \right]  \right\}  ^{1/2}\right\}  ^{-1}. \label{2DBEVz}%
\end{align}
This relation between $z$\ and $T$\ is plotted in figure 3.

\begin{figure}[ptb]
\begin{center}
\includegraphics[height=2.5in]
{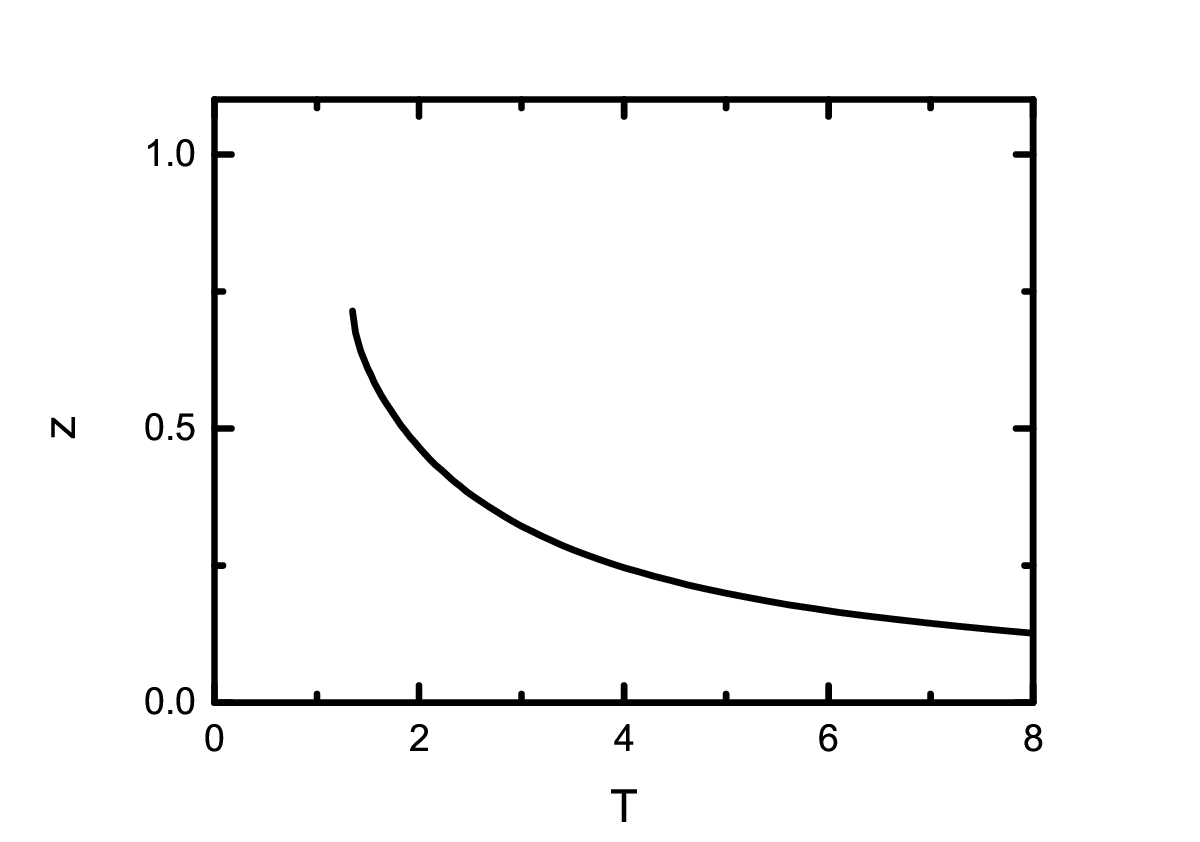}
\end{center}
\caption{The relation between $z$ and $T$ for two-dimensional hard-sphere Bose
gases in an isochoric process.}%
\label{fig3}%
\end{figure}

Substituting the fugacity $z$ given by Eq. (\ref{2DBEVz}) into Eq.
(\ref{2DBEP}) gives the equation of state.

\subsection{Two-dimensional hard-sphere Fermi gases}

The equation of state for two-dimensional weakly interacting hard-sphere Fermi
gases, up to first order of $a/\lambda$, is \cite{OurEPL}%

\begin{align}
\frac{P}{kT}  &  =\left(  2j+1\right)  \frac{1}{\lambda^{2}}\left[
f_{2}\left(  z\right)  -j\pi\frac{a}{\lambda}f_{1}\left(  z\right)
f_{3/2}\left(  z\right)  \right]  ,\label{2DFDP}\\
n  &  =\left(  2j+1\right)  \frac{1}{\lambda^{2}}\left\{  f_{1}\left(
z\right)  -j\pi\frac{a}{\lambda}\left[  f_{0}\left(  z\right)  f_{3/2}\left(
z\right)  +f_{1/2}\left(  z\right)  f_{1}\left(  z\right)  \right]  \right\}
. \label{N2DF}%
\end{align}
The fugacity, determined by Eq. (\ref{N2DF}), is a zero of the complex
function%
\begin{equation}
\Psi_{4}\left(  \zeta\right)  =\left(  2j+1\right)  \frac{1}{n\lambda^{2}%
}\left\{  f_{1}\left(  \zeta\right)  -j\pi\frac{a}{\lambda}\left[
f_{0}\left(  \zeta\right)  f_{3/2}\left(  \zeta\right)  +f_{1/2}\left(
\zeta\right)  f_{1}\left(  \zeta\right)  \right]  \right\}  -1
\end{equation}
on the real axis.

Like that in the case of three-dimensional hard-sphere Fermi gases, the
singularities of $\Psi_{4}\left(  \zeta\right)  $ form a line from $-1$ to
$-\infty$, and $\Psi_{4}\left(  \zeta\right)  $ has two real zeros and has no
isolated singularities. Thus, $\Psi_{4}\left(  \zeta\right)  $ can be
expressed as%

\begin{equation}
\Psi_{4}\left(  \zeta\right)  =\eta_{4}\left(  \zeta-z\right)  \left(
\zeta-\omega\right)  \Phi_{4}\left(  \zeta\right)  ,
\end{equation}
where $\Phi_{4}\left(  \zeta\right)  $ has no zeros and has the same
singularities as those of $\Psi_{4}\left(  \zeta\right)  $.

The fundamental solution of the homogeneous Riemann-Hilbert problem is chosen
to satisfy Eq. (\ref{ch}). Near the unique end different from infinity point,
$\zeta=-1$, on the boundary, we have%
\begin{equation}
\left.  \Psi_{4}\left(  \zeta\right)  \right\vert _{\zeta\rightarrow-1}%
\sim\frac{1}{\zeta+1},
\end{equation}
i.e., $\Psi_{4}\left(  \zeta\right)  $, and so $\Phi_{4}\left(  \zeta\right)
$, linearly diverges near $\zeta=-1$, which means $\Phi_{4}\left(
\zeta\right)  $ is not a fundamental solution. According to Eq. (\ref{II7}),
we choose $\beta=1$ and express $\Phi_{4}\left(  \zeta\right)  $ in terms of
the fundamental solution $\varphi_{4}\left(  \zeta\right)  $:%
\begin{equation}
\Phi_{4}\left(  \zeta\right)  =\frac{\varphi_{4}\left(  \zeta\right)  }%
{\zeta+1}.
\end{equation}

From the jump on the boundary $\left[  -1,-\infty\right)  $,%
\begin{equation}
G_{4}\left(  x\right)  =\frac{\varphi_{4}^{+}\left(  x\right)  }{\varphi
_{4}^{-}\left(  x\right)  }=\frac{\Psi_{4}^{+}\left(  x\right)  }{\Psi_{4}%
^{-}\left(  x\right)  },\text{ }x\in\left[  -1,-\infty\right)  ,
\end{equation}
we can determine the fundamental solution $\varphi_{4}\left(  \zeta\right)  $.
From Eqs. (\ref{FDI}) and (\ref{PFDI}), one reaches%
\begin{align}
\Psi_{4}^{\pm}\left(  x\right)   &  =\left(  2j+1\right)  \frac{1}%
{n\lambda^{2}}\left\{  \mathfrak{f}_{1}\left(  x\right)  -j\pi\frac{a}%
{\lambda}\left[  f_{0}\left(  x\right)  \mathfrak{f}_{3/2}\left(  x\right)
+\mathfrak{f}_{1/2}\left(  x\right)  \mathfrak{f}_{1}\left(  x\right)
-\frac{\pi^{3/2}}{\sqrt{\ln\left(  -x\right)  }}\right]  \right\}
-1\nonumber\\
&  \mp i\left(  2j+1\right)  \frac{1}{n\lambda^{2}}\pi\left\{  1-j\sqrt{\pi
}\frac{a}{\lambda}\left[  2\sqrt{\ln\left(  -x\right)  }f_{0}\left(  x\right)
+\sqrt{\pi}\mathfrak{f}_{1/2}\left(  x\right)  +\frac{1}{\sqrt{\ln\left(
-x\right)  }}\mathfrak{f}_{1}\left(  x\right)  \right]  \right\}  .
\end{align}
Notice that when $\sigma=0$, $f_{0}\left(  \zeta\right)  =\zeta/\left(
1+\zeta\right)  $ has only one isolated singularity, $\zeta=-1$. Since
$\left[  \Psi_{4}^{+}\left(  x\right)  \right]  ^{\ast}=\Psi_{4}^{-}\left(
x\right)  $, the jump
\begin{equation}
G_{4}\left(  x\right)  =\exp\left[  i2\arg\Psi_{4}^{+}\left(  x\right)
\right]  ,
\end{equation}
where $\arg\Psi_{4}^{+}\left(  x\right)  =\operatorname{arccot}\left[
\operatorname{Re}\Psi_{4}^{+}\left(  x\right)  /\operatorname{Im}\Psi_{4}%
^{+}\left(  x\right)  \right]  $. In this case, there is only one end point
different from infinity, $\zeta=-1$, so Eq. (\ref{II8}) gives the fundamental
solution%
\begin{equation}
\varphi_{4}\left(  \zeta\right)  =e^{\gamma_{4}\left(  \zeta\right)  }\left(
\zeta+1\right)  ^{\lambda_{4}},
\end{equation}
where%
\begin{equation}
\gamma_{4}\left(  \zeta\right)  =\frac{1}{\pi}%
%TCIMACRO{\dint \nolimits_{-1}^{-\infty}}%
%BeginExpansion
{\displaystyle\int\nolimits_{-1}^{-\infty}}
%EndExpansion
dx\frac{\arg\Psi_{4}^{+}\left(  x\right)  }{x-\zeta}.
\end{equation}
The fundamental solution is chosen to satisfy Eq. (\ref{ch}). Choosing
$\arg\Psi_{4}^{+}\left(  -\infty\right)  =0$, then $\arg\Psi_{4}^{+}\left(
-1\right)  =-\pi$, from the condition Eq. (\ref{II9}), we have%
\begin{equation}
-\frac{1}{\pi}\arg\Psi_{4}^{+}\left(  -1\right)  +\lambda_{4}=0,
\end{equation}
which gives $\lambda_{4}=-1$. Finally,%
\begin{equation}
\Psi_{4}\left(  \zeta\right)  =\eta_{4}\frac{\left(  \zeta-z\right)  \left(
\zeta-\omega\right)  }{\left(  \zeta+1\right)  ^{2}}e^{\gamma_{4}\left(
\zeta\right)  }. \label{zero4}%
\end{equation}

By setting $\zeta=0$, Eq. (\ref{zero4}) and its first- and second-order
derivatives give three equations to determine $z$, $\omega$ and the constant
$\eta_{4}$, whose solution is%
\begin{align}
z  &  =2\frac{n\lambda^{2}}{2j+1}\left\{  1+\frac{n\lambda^{2}}{2j+1}\left[
\gamma_{4}^{\prime}\left(  0\right)  -2\right]  +\left\{  1-2\frac
{n\lambda^{2}}{2j+1}\left[  \gamma_{4}^{\prime}\left(  0\right)  -1+4\pi
j\frac{a}{\lambda}\right]  \right.  \right. \nonumber\\
&  \left.  \left.  +\frac{\left(  n\lambda^{2}\right)  ^{2}}{\left(
2j+1\right)  ^{2}}\left[  4\gamma_{4}^{\prime}\left(  0\right)  -\gamma
_{4}^{\prime}\left(  0\right)  ^{2}+2\gamma_{4}^{\prime\prime}\left(
0\right)  \right]  \right\}  ^{1/2}\right\}  ^{-1}. \label{2DFDVz}%
\end{align}
This relation between $z$\ and $T$\ is plotted in figure 4.

\begin{figure}[ptb]
\begin{center}
\includegraphics[height=2.5in]
{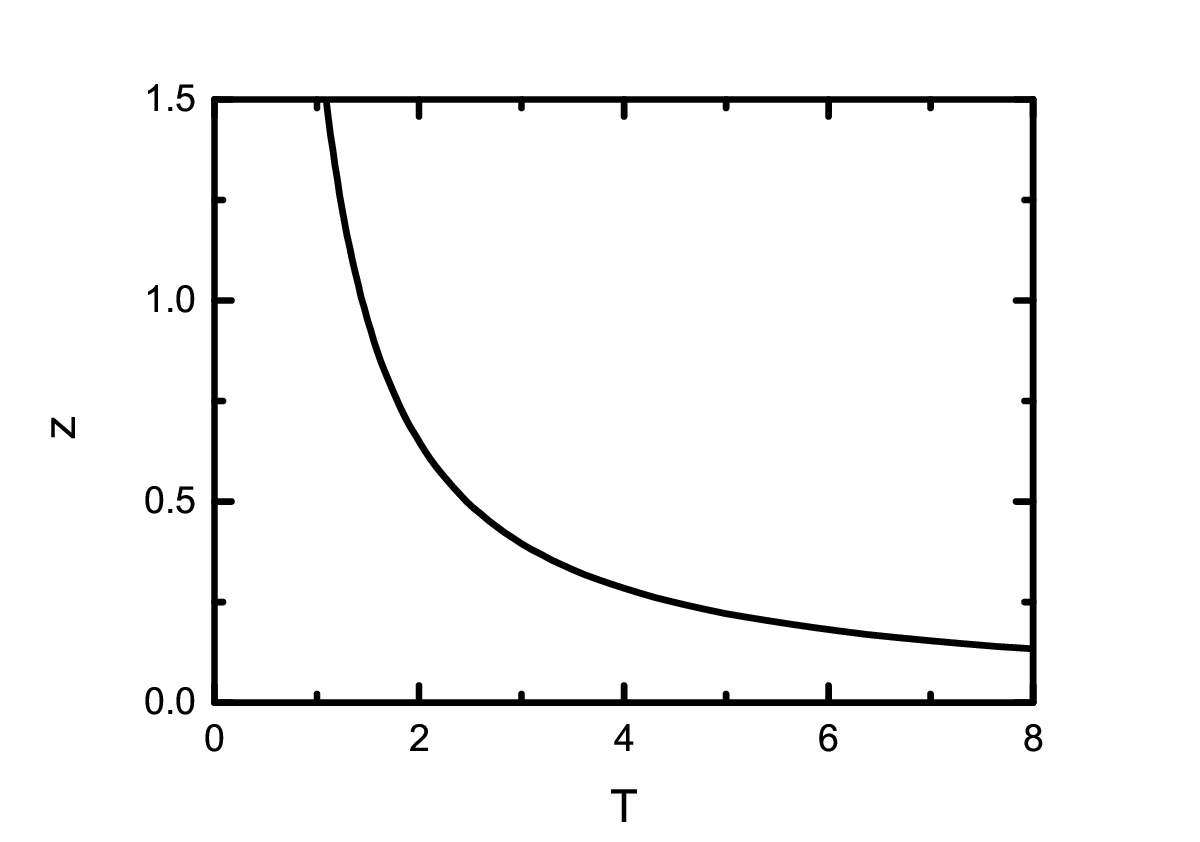}
\end{center}
\caption{The relation between $z$ and $T$ for two-dimensional hard-sphere
Fermi gases in an isochoric process.}%
\label{fig4}%
\end{figure}

Substituting the fugacity $z$ given by Eq. (\ref{2DFDVz}) into Eq.
(\ref{2DFDP}) gives the equation of state.

\section{The explicit expression for the fugacity: isobaric processes\label{V}%
}

In the above section, we calculate the explicit expression for the fugacity in
isochoric processes. In this section, we calculate the explicit expression for
the fugacity for hard-sphere quantum gases in isobaric processes. Isobaric
processes are of great significance to the problem of phase transition. In an
isobaric process, the fugacity is determined by Eq. (\ref{I1}).

\subsection{Three-dimensional hard-sphere Bose gases}

The equation of state for hard-sphere Bose gases is given by Eqs. (\ref{3DB1})
and (\ref{3DB2}). In isobaric processes, from Eq. (\ref{3DB1}), the fugacity
$z$ is a zero of the complex function%
\begin{equation}
\Psi_{5}\left(  \zeta\right)  =\left(  2j+1\right)  \frac{kT}{P\lambda^{3}%
}\left[  g_{5/2}\left(  \zeta\right)  -2\left(  j+1\right)  \frac{a}{\lambda
}g_{3/2}^{2}\left(  \zeta\right)  \right]  -1
\end{equation}
on the real axis.

There are two different cases corresponding to different values of parameters:
$\Psi_{5}\left(  \zeta\right)  $ has two zeros and $\Psi_{5}\left(
\zeta\right)  $ has one zero.

\subsubsection{The case of two zeros}

The boundary of the analytic region of $\Psi_{5}\left(  \zeta\right)  $ is the
line $\left[  1,\infty\right)  $, and $\Psi_{5}\left(  \zeta\right)  $ can be
written as%
\begin{equation}
\Psi_{5}\left(  \zeta\right)  =\eta_{5}\left(  \zeta-z\right)  \left(
\zeta-\omega\right)  \Phi_{5}\left(  \zeta\right)  .
\end{equation}
$\Phi_{5}\left(  \zeta\right)  $ has no zeros and has the same singularities
as $\Psi_{5}\left(  \zeta\right)  $. Since $\Psi_{5}\left(  \zeta\right)  $ is
convergent at the unique end different from infinity, $\zeta=1$, of the
boundary, $\Phi_{5}\left(  \zeta\right)  =\varphi_{5}\left(  \zeta\right)  $
is just a fundamental solution.

At the two sides of the boundary $\left[  1,\infty\right)  $, by Eq.
(\ref{BEICP}), we have%
\begin{align}
\Psi_{5}^{\pm}\left(  x\right)   &  =\left(  2j+1\right)  \frac{kT}%
{P\lambda^{3}}\left\{  \mathfrak{g}_{5/2}\left(  x\right)  -2\left(
j+1\right)  \frac{a}{\lambda}\left[  \mathfrak{g}_{3/2}^{2}\left(  x\right)
-4\pi\ln x\right]  \right\}  -1\nonumber\\
&  \pm i\left(  2j+1\right)  \frac{kT}{P\lambda^{3}}4\sqrt{\pi}\sqrt{\ln
x}\left[  \frac{1}{3}\ln x-2\left(  j+1\right)  \frac{a}{\lambda}%
\mathfrak{g}_{3/2}\left(  x\right)  \right]  .
\end{align}
The jump is then%
\begin{equation}
G_{5}\left(  x\right)  =\frac{\varphi_{5}^{+}\left(  x\right)  }{\varphi
_{5}^{-}\left(  x\right)  }=\frac{\Psi_{5}^{+}\left(  x\right)  }{\Psi_{5}%
^{-}\left(  x\right)  }=\exp\left[  i2\arg\Psi_{5}^{+}\left(  x\right)
\right]  ,\text{ }x\in\left[  1,\infty\right)  ,
\end{equation}
where $\arg\Psi_{5}^{+}\left(  x\right)  =\operatorname{arccot}\left[
\operatorname{Re}\Psi_{5}^{+}\left(  x\right)  /\operatorname{Im}\Psi_{5}%
^{+}\left(  x\right)  \right]  $. The fundamental solution is consequently%
\begin{equation}
\varphi_{5}\left(  \zeta\right)  =e^{\gamma_{5}\left(  \zeta\right)  }\left(
\zeta-1\right)  ^{\lambda_{5}},
\end{equation}
where%
\begin{equation}
\gamma_{5}\left(  \zeta\right)  =\frac{1}{\pi}%
%TCIMACRO{\dint \nolimits_{1}^{\infty}}%
%BeginExpansion
{\displaystyle\int\nolimits_{1}^{\infty}}
%EndExpansion
dx\frac{\arg\Psi_{5}^{+}\left(  x\right)  }{x-\zeta}.
\end{equation}
The fundamental solution is chosen to satisfy Eq. (\ref{ch}). We choose
$\arg\Psi_{5}^{+}\left(  \infty\right)  =0$, then $\arg\Psi_{5}^{+}\left(
1\right)  =-2\pi$, thus the condition (\ref{II9})%
\begin{equation}
-\frac{1}{\pi}\arg\Psi_{5}^{+}\left(  1\right)  +\lambda_{5}=0
\end{equation}
gives $\lambda_{5}=-2$, and so%

\begin{equation}
\Psi_{5}\left(  \zeta\right)  =\eta_{5}\frac{\left(  \zeta-z\right)  \left(
\zeta-\omega\right)  }{\left(  \zeta-1\right)  ^{2}}e^{\gamma_{5}\left(
\zeta\right)  }. \label{zero5}%
\end{equation}

By setting $\zeta=0$, Eq. (\ref{zero5}) and its first- and second-order
derivatives give three equations to determine the zeros $z$, $\omega$ and the
constant $\eta_{5}$, and the solution is%
\begin{align}
z  &  =2\frac{P\lambda^{3}}{\left(  2j+1\right)  kT}\left\{  1+\frac
{P\lambda^{3}}{\left(  2j+1\right)  kT}\left[  \gamma_{5}^{\prime}\left(
0\right)  +2\right]  +\left\{  1-2\frac{P\lambda^{3}}{\left(  2j+1\right)
kT}\left[  \gamma_{5}^{\prime}\left(  0\right)  +2-\frac{\sqrt{2}}{4}+4\left(
j+1\right)  \pi\frac{a}{\lambda}\right]  \right.  \right. \nonumber\\
&  \left.  \left.  -\left(  \frac{P\lambda^{3}}{\left(  2j+1\right)
kT}\right)  ^{2}\left[  4\gamma_{5}^{\prime}\left(  0\right)  +\gamma
_{5}^{\prime}\left(  0\right)  ^{2}-2\gamma_{5}^{\prime\prime}\left(
0\right)  \right]  \right\}  ^{1/2}\right\}  ^{-1}. \label{ztwo0}%
\end{align}

\subsubsection{The case of one zero}

When $\Psi_{5}\left(  \zeta\right)  $ has only one zero, we define%
\begin{equation}
\Psi_{5}\left(  \zeta\right)  =\eta_{5}\left(  \zeta-z\right)  \Phi_{5}\left(
\zeta\right)  .
\end{equation}
We still have $\Phi_{5}\left(  \zeta\right)  =\varphi_{5}\left(  \zeta\right)
$ and%
\begin{equation}
\varphi_{5}\left(  \zeta\right)  =e^{\gamma_{5}\left(  \zeta\right)  }\left(
\zeta-1\right)  ^{\lambda_{5}}.
\end{equation}
When choosing $\arg\Psi_{5}^{+}\left(  \infty\right)  =0$, we have $\arg
\Psi_{5}^{+}\left(  1\right)  =-\pi$. Then the condition (\ref{II9})%
\begin{equation}
-\frac{1}{\pi}\arg\Psi_{5}^{+}\left(  1\right)  +\lambda_{5}=0
\end{equation}
gives $\lambda_{5}=-1$, so%

\begin{equation}
\Psi_{5}\left(  \zeta\right)  =\eta_{5}\frac{\zeta-z}{\zeta-1}e^{\gamma
_{5}\left(  \zeta\right)  }.
\end{equation}
We have%
\begin{equation}
z=\left[  \left(  2j+1\right)  \frac{kT}{P\lambda^{3}}+\gamma_{5}{}^{\prime
}\left(  0\right)  +1\right]  ^{-1}. \label{3DBEPz}%
\end{equation}
This relation between $z$\ and $T$\ is plotted in figure 5.

\begin{figure}[ptb]
\begin{center}
\includegraphics[height=2.5in]
{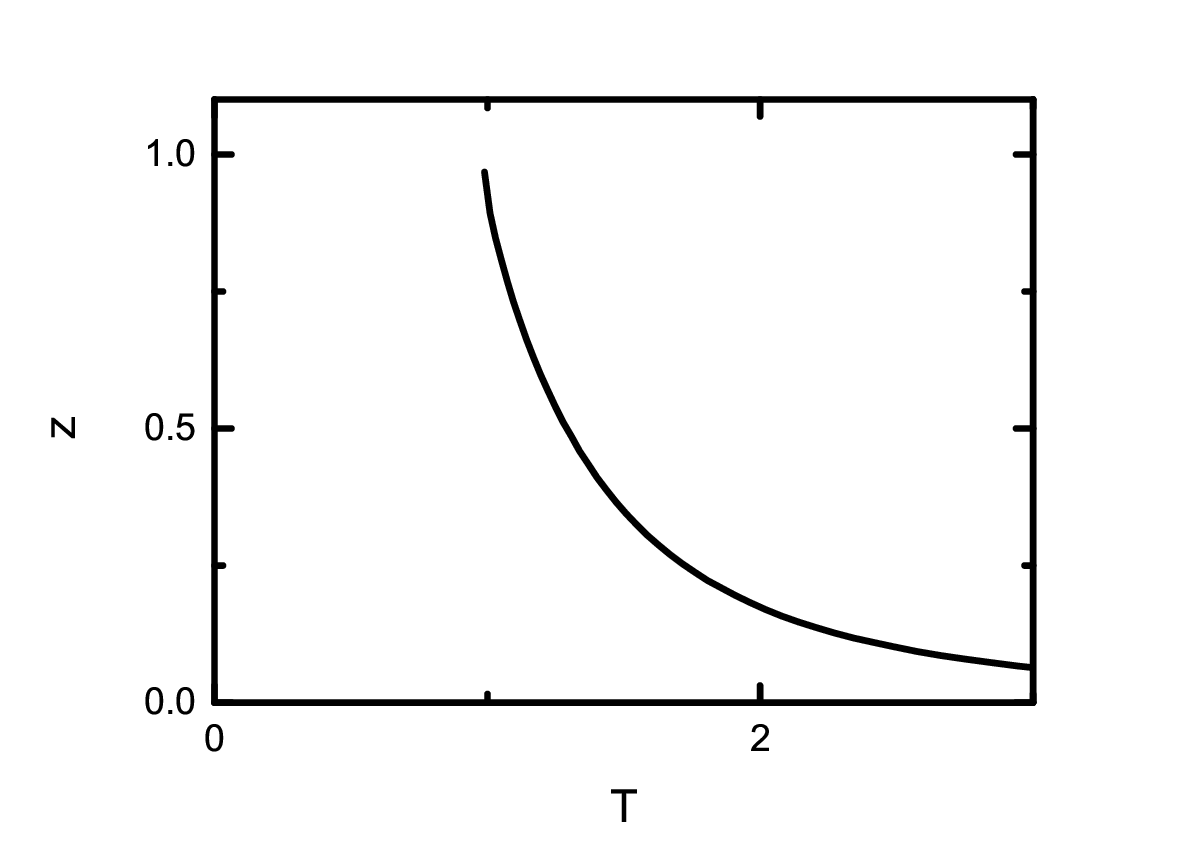}
\end{center}
\caption{The relation between $z$ and $T$ for three-dimensional hard-sphere
Bose gases in an isobaric process.}%
\label{fig5}%
\end{figure}

Substituting the fugacity $z$ given by Eq. (\ref{ztwo0}) or (\ref{3DBEPz})
into Eq. (\ref{3DB2}) gives the equation of state, respectively.

\subsection{Three-dimensional hard-sphere Fermi gases}

Similarly, for the case of hard-sphere Fermi gases, according to Eq.
(\ref{3DFDP}), we define%
\begin{equation}
\Psi_{6}\left(  \zeta\right)  =\left(  2j+1\right)  \frac{kT}{P\lambda^{3}%
}\left[  f_{5/2}\left(  \zeta\right)  -2j\frac{a}{\lambda}f_{3/2}^{2}\left(
\zeta\right)  \right]  -1,
\end{equation}
which has two zeros on the real axis and the boundary of its analytic region
is $\left[  -1,-\infty\right)  $. $\Psi_{6}\left(  \zeta\right)  $ can be
expressed as%
\begin{equation}
\Psi_{6}\left(  \zeta\right)  =\eta_{6}\left(  \zeta-z\right)  \left(
\zeta-\omega\right)  \Phi_{6}\left(  \zeta\right)  .
\end{equation}
As $\Psi_{6}\left(  \zeta\right)  $ converges at the unique end different from
infinity of the boundary, $\zeta=-1$, $\Phi_{6}\left(  \zeta\right)
=\varphi_{6}\left(  \zeta\right)  $ is a fundamental solution.

At the two sides of the boundary $\left[  -1,-\infty\right)  $,%
\begin{align}
\Psi_{6}^{\pm}\left(  x\right)   &  =\left(  2j+1\right)  \frac{kT}%
{P\lambda^{3}}\left\{  \mathfrak{f}_{5/2}\left(  x\right)  -2j\frac{a}%
{\lambda}\left[  \mathfrak{f}_{3/2}^{2}\left(  x\right)  -4\pi\ln\left(
-x\right)  \right]  \right\}  -1\nonumber\\
&  \mp i4\sqrt{\pi}\left(  2j+1\right)  \frac{kT}{P\lambda^{3}}\sqrt
{\ln\left(  -x\right)  }\left[  \frac{1}{3}\ln\left(  -x\right)  -2j\frac
{a}{\lambda}\mathfrak{f}_{3/2}\left(  x\right)  \right]  ,
\end{align}
and the jump is%
\begin{equation}
G_{6}\left(  x\right)  =\frac{\varphi_{6}^{+}\left(  x\right)  }{\varphi
_{6}^{-}\left(  x\right)  }=\frac{\Psi_{6}^{+}\left(  x\right)  }{\Psi_{6}%
^{-}\left(  x\right)  }=\exp\left[  i2\arg\Psi_{6}^{+}\left(  x\right)
\right]  ,\text{ }x\in\left[  -1,-\infty\right)  ,
\end{equation}
where $\arg\Psi_{6}^{+}\left(  x\right)  =\operatorname{arccot}\left[
\operatorname{Re}\Psi_{6}^{+}\left(  x\right)  /\operatorname{Im}\Psi_{6}%
^{+}\left(  x\right)  \right]  $. Thus, the fundamental solution is%
\begin{equation}
\varphi_{6}\left(  \zeta\right)  =e^{\gamma_{6}\left(  \zeta\right)  }\left(
\zeta+1\right)  ^{\lambda_{6}},
\end{equation}
where%
\begin{equation}
\gamma_{6}\left(  \zeta\right)  =\frac{1}{\pi}%
%TCIMACRO{\dint \nolimits_{-1}^{-\infty}}%
%BeginExpansion
{\displaystyle\int\nolimits_{-1}^{-\infty}}
%EndExpansion
dx\frac{\arg\Psi_{6}^{+}\left(  x\right)  }{x-\zeta}.
\end{equation}
The fundamental solution is chosen to satisfy Eq. (\ref{ch}). Choosing
$\arg\Psi_{6}^{+}\left(  -\infty\right)  =0$ gives $\arg\Psi_{6}^{+}\left(
-1\right)  =-2\pi$. Then, the condition (\ref{II9})%
\begin{equation}
-\frac{1}{\pi}\arg\Psi_{6}^{+}\left(  -1\right)  +\lambda_{6}=0
\end{equation}
gives $\lambda_{6}=-2$. Consequently,%

\begin{equation}
\Psi_{6}\left(  \zeta\right)  =\eta_{6}\frac{\left(  \zeta-z\right)  \left(
\zeta-\omega\right)  }{\left(  \zeta+1\right)  ^{2}}e^{\gamma_{6}\left(
\zeta\right)  }. \label{zero6}%
\end{equation}

By setting $\zeta=0$, Eq. (\ref{zero6}) and its first- and second-order
derivatives give three equations for $z$, $\omega$ and the constant $\eta_{6}%
$, and the solution of the fugacity is%
\begin{align}
z  &  =2\frac{P\lambda^{3}}{\left(  2j+1\right)  kT}\left\{  1+\frac
{P\lambda^{3}}{\left(  2j+1\right)  kT}\left[  \gamma_{6}^{\prime}\left(
0\right)  -2\right]  +\left\{  1-2\frac{P\lambda^{3}}{\left(  2j+1\right)
kT}\left[  \gamma_{6}^{\prime}\left(  0\right)  -2+\frac{\sqrt{2}}{4}%
+4j\frac{a}{\lambda}\right]  \right.  \right. \nonumber\\
&  \left.  \left.  +\left(  \frac{P\lambda^{3}}{\left(  2j+1\right)
kT}\right)  ^{2}\left[  4\gamma_{6}^{\prime}\left(  0\right)  -\gamma
_{6}^{\prime}\left(  0\right)  ^{2}+2\gamma_{6}^{\prime\prime}\left(
0\right)  \right]  \right\}  ^{1/2}\right\}  ^{-1}. \label{3DFDPz}%
\end{align}
This relation between $z$\ and $T$\ is plotted in figure 6.

\begin{figure}[ptb]
\begin{center}
\includegraphics[height=2.5in]
{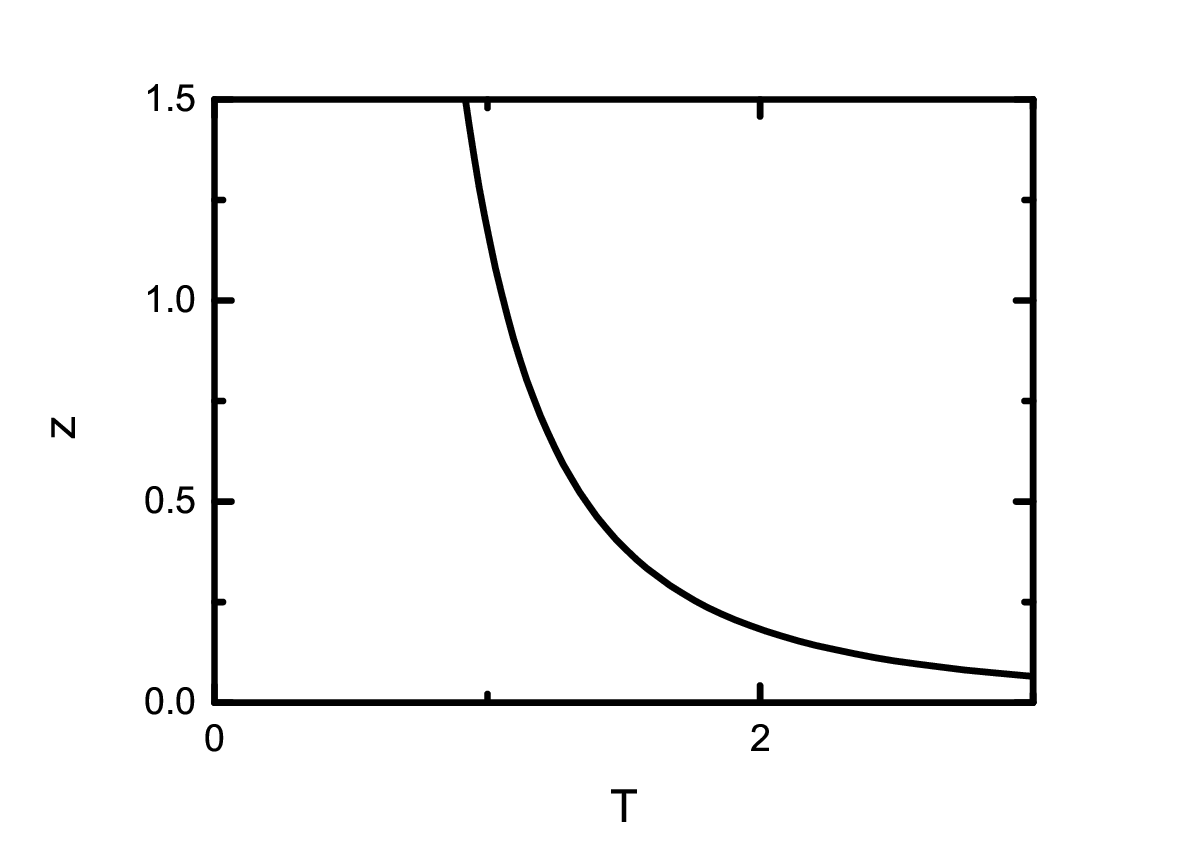}
\end{center}
\caption{The relation between $z$ and $T$ for three-dimensional hard-sphere
Fermi gases in an isobaric process.}%
\label{fig6}%
\end{figure}

Substituting the fugacity $z$ given by Eq. (\ref{3DFDPz}) into Eq.
(\ref{3DFDn}) gives the equation of state.

\subsection{Two-dimensional hard-sphere Bose gases}

In an isobaric process, the fugacity for two-dimensional hard-sphere Bose
gases is given by Eq. (\ref{2DBEP}). The fugacity $z$ is a zero of the
function%
\begin{equation}
\Psi_{7}\left(  \zeta\right)  =\left(  2j+1\right)  \frac{kT}{P\lambda^{2}%
}\left[  g_{2}\left(  \zeta\right)  -\left(  j+1\right)  \pi\frac{a}{\lambda
}g_{1}\left(  \zeta\right)  g_{3/2}\left(  \zeta\right)  \right]  -1
\end{equation}
on the real axis. $\Psi_{7}\left(  \zeta\right)  $ has two zeros, $z$ and
$\omega$, and the boundary of its analytic region is $\left[  1,\infty\right)
$. Express $\Psi_{7}\left(  \zeta\right)  $ as%
\begin{equation}
\Psi_{7}\left(  \zeta\right)  =\eta_{7}\left(  \zeta-z\right)  \left(
\zeta-\omega\right)  \Phi_{7}\left(  \zeta\right)  .
\end{equation}
The function $\Phi_{7}\left(  \zeta\right)  $ has no zeros in the complex
$\zeta$-plane. At the unique end different from infinity of the boundary,
$\zeta=1$, $\Psi_{7}\left(  \zeta\right)  $ is logarithmically divergent:%
\begin{equation}
\left.  \Psi_{7}\left(  \zeta\right)  \right\vert _{\zeta\rightarrow1}\sim
\ln\left(  \zeta-1\right)  ,
\end{equation}
i.e., the degree of divergence is less than $1$, so $\Phi_{7}\left(
\zeta\right)  =\varphi_{7}\left(  \zeta\right)  $ is a fundamental solution.

The values of $\Psi_{7}\left(  \zeta\right)  $ at the two sides of the
boundary $\left[  1,\infty\right)  $ are%
\begin{align}
\Psi_{7}^{\pm}\left(  x\right)   &  =\left(  2j+1\right)  \frac{kT}%
{P\lambda^{2}}\left\{  \mathfrak{g}_{2}\left(  x\right)  -\left(  j+1\right)
\pi\frac{a}{\lambda}\left[  \mathfrak{g}_{1}\left(  x\right)  \mathfrak{g}%
_{3/2}\left(  x\right)  -2\pi^{3/2}\sqrt{\ln x}\right]  \right\}
-1\nonumber\\
&  \pm i\left(  2j+1\right)  \frac{kT}{P\lambda^{2}}\pi\left\{  \ln x-\left(
j+1\right)  \frac{a}{\lambda}\left[  2\sqrt{\pi}\sqrt{\ln x}\mathfrak{g}%
_{1}\left(  x\right)  +\pi\mathfrak{g}_{3/2}\left(  x\right)  \right]
\right\}  .
\end{align}
The jump is therefore%
\begin{equation}
G_{7}\left(  x\right)  =\frac{\varphi_{7}^{+}\left(  x\right)  }{\varphi
_{7}^{-}\left(  x\right)  }=\frac{\Psi_{7}^{+}\left(  x\right)  }{\Psi_{7}%
^{-}\left(  x\right)  }=\exp\left[  i2\arg\Psi_{7}^{+}\left(  x\right)
\right]  ,\text{ }x\in\left[  1,\infty\right)  ,
\end{equation}
where $\arg\Psi_{7}^{+}\left(  x\right)  =\operatorname{arccot}\left[
\operatorname{Re}\Psi_{7}^{+}\left(  x\right)  /\operatorname{Im}\Psi_{7}%
^{+}\left(  x\right)  \right]  $. The fundamental solution is then%
\begin{equation}
\varphi_{7}\left(  \zeta\right)  =e^{\gamma_{7}\left(  \zeta\right)  }\left(
\zeta-1\right)  ^{\lambda_{7}},
\end{equation}
where%
\begin{equation}
\gamma_{7}\left(  \zeta\right)  =\frac{1}{\pi}%
%TCIMACRO{\dint \nolimits_{1}^{\infty}}%
%BeginExpansion
{\displaystyle\int\nolimits_{1}^{\infty}}
%EndExpansion
dx\frac{\arg\Psi_{7}^{+}\left(  x\right)  }{x-\zeta}.
\end{equation}
The fundamental solution is chosen to satisfy Eq. (\ref{ch}). When choosing
$\arg\Psi_{7}^{+}\left(  \infty\right)  =0$, $\arg\Psi_{7}^{+}\left(
1\right)  =-2\pi$, the condition (\ref{II9})%
\begin{equation}
-\frac{1}{\pi}\arg\Psi_{7}^{+}\left(  1\right)  +\lambda_{7}=0
\end{equation}
gives $\lambda_{7}=-2$. Therefore,%
\begin{equation}
\Psi_{7}\left(  \zeta\right)  =\eta_{7}\frac{\left(  \zeta-z\right)  \left(
\zeta-\omega\right)  }{\left(  \zeta-1\right)  ^{2}}e^{\gamma_{7}\left(
\zeta\right)  }. \label{zero7}%
\end{equation}
By setting $\zeta=0$, Eq. (\ref{zero7}) and its first- and second-order
derivatives give three equations for the zeros $z$, $\omega$ and the constant
$\eta_{7}$, which give the solution of the fugacity%
\begin{align}
z  &  =2\frac{P\lambda^{2}}{\left(  2j+1\right)  kT}\left\{  1+\frac
{P\lambda^{2}}{\left(  2j+1\right)  kT}\left[  \gamma_{7}^{\prime}\left(
0\right)  +2\right]  +\left\{  1-2\frac{P\lambda^{2}}{\left(  2j+1\right)
kT}\left[  \gamma_{7}^{\prime}\left(  0\right)  +\frac{3}{2}+2\left(
j+1\right)  \pi\frac{a}{\lambda}\right]  \right.  \right. \nonumber\\
&  \left.  \left.  -\left(  \frac{P\lambda^{2}}{\left(  2j+1\right)
kT}\right)  ^{2}\left[  4\gamma_{7}^{\prime}\left(  0\right)  +\gamma
_{7}^{\prime}\left(  0\right)  ^{2}-2\gamma_{7}^{\prime\prime}\left(
0\right)  \right]  \right\}  ^{1/2}\right\}  ^{-1}. \label{2DBEPz}%
\end{align}
This relation between $z$\ and $T$\ is plotted in figure 7.

\begin{figure}[ptb]
\begin{center}
\includegraphics[height=2.5in]
{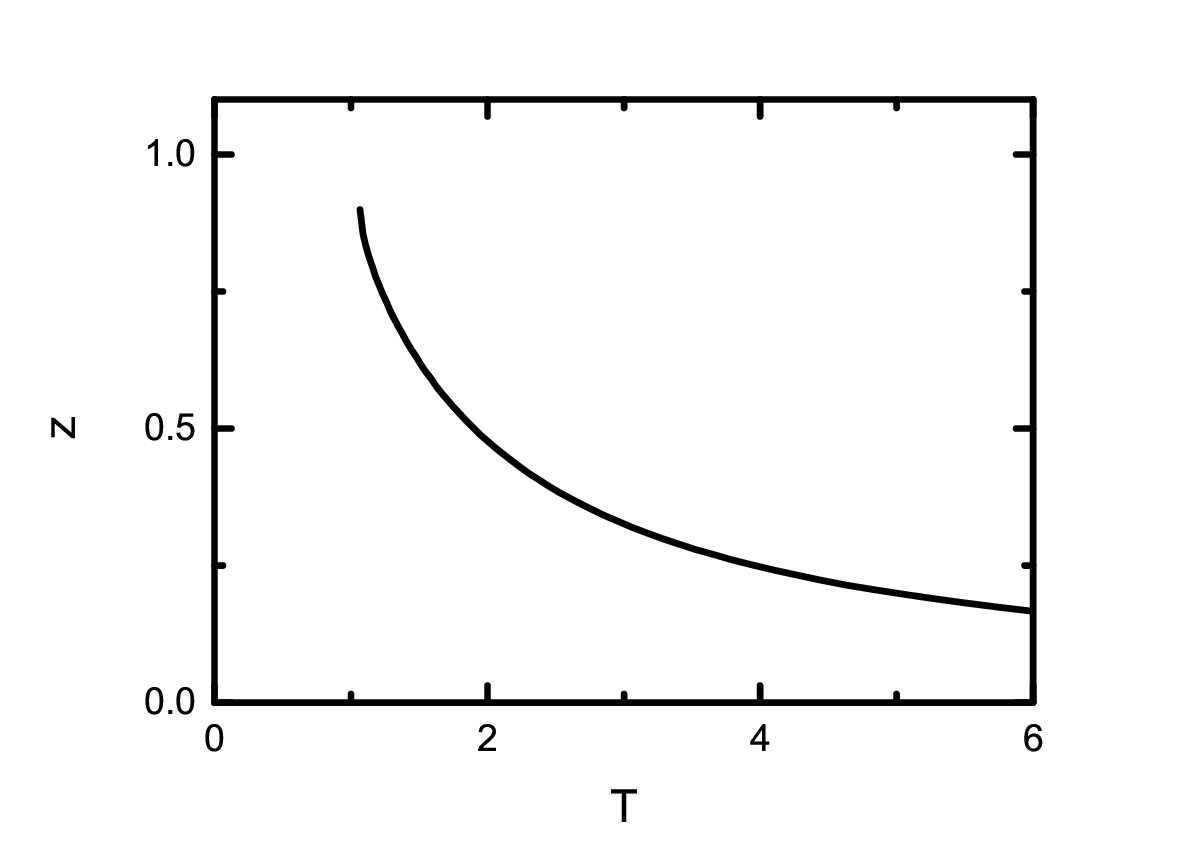}
\end{center}
\caption{The relation between $z$ and $T$ for two-dimensional hard-sphere Bose
gases in an isobaric process.}%
\label{fig7}%
\end{figure}

Substituting the fugacity $z$ given by Eq. (\ref{2DBEPz}) into Eq.
(\ref{2DBEn}) gives the equation of state.

\subsection{Two-dimensional hard-sphere Fermi gases}

The fugacity for two-dimensional hard-sphere Fermi gases is given by Eq.
(\ref{2DFDP}). The function%

\begin{equation}
\Psi_{8}\left(  \zeta\right)  =\left(  2j+1\right)  \frac{kT}{P\lambda^{2}%
}\left[  f_{2}\left(  \zeta\right)  -j\pi\frac{a}{\lambda}f_{1}\left(
\zeta\right)  f_{3/2}\left(  \zeta\right)  \right]  -1
\end{equation}
has two zeros on the real axis. The boundary of the analytic region of
$\Psi_{8}\left(  \zeta\right)  $ is $\left[  -1,-\infty\right)  $. Express
$\Psi_{8}\left(  \zeta\right)  $ as%
\begin{equation}
\Psi_{8}\left(  \zeta\right)  =\eta_{8}\left(  \zeta-z\right)  \left(
\zeta-\omega\right)  \Phi_{8}\left(  \zeta\right)  .
\end{equation}
$\Phi_{8}\left(  \zeta\right)  $ has no zeros in the complex $\zeta$-plane. At
the unique end different from infinity of the boundary, $\zeta=-1$, $\Psi
_{8}\left(  \zeta\right)  $ is logarithmically divergent:
\begin{equation}
\left.  \Psi_{8}\left(  \zeta\right)  \right\vert _{\zeta\rightarrow-1}\sim
\ln\left(  \zeta+1\right)  ,
\end{equation}
i.e., the degree of divergence is less than $1$, so $\Phi_{8}\left(
\zeta\right)  =\varphi_{8}\left(  \zeta\right)  $ is a fundamental solution.

The values of $\Psi_{8}\left(  \zeta\right)  $ at the two sides of the
boundary $\left[  -1,-\infty\right)  $ are%
\begin{align}
\Psi_{8}^{\pm}\left(  x\right)   &  =\left(  2j+1\right)  \frac{kT}%
{P\lambda^{2}}\left\{  \mathfrak{f}_{2}\left(  x\right)  -j\pi\frac{a}%
{\lambda}\left[  \mathfrak{f}_{1}\left(  x\right)  \mathfrak{f}_{3/2}\left(
x\right)  -2\pi^{3/2}\sqrt{\ln\left(  -x\right)  }\right]  \right\}
-1\nonumber\\
&  \mp i\left(  2j+1\right)  \frac{kT}{P\lambda^{2}}\pi\left\{  \ln\left(
-x\right)  -j\frac{a}{\lambda}\left[  2\sqrt{\pi}\sqrt{\ln\left(  -x\right)
}\mathfrak{f}_{1}\left(  x\right)  +\pi\mathfrak{f}_{3/2}\left(  x\right)
\right]  \right\}  .
\end{align}
The jump is therefore%
\begin{equation}
G_{8}\left(  x\right)  =\frac{\varphi_{8}^{+}\left(  x\right)  }{\varphi
_{8}^{-}\left(  x\right)  }=\frac{\Psi_{8}^{+}\left(  x\right)  }{\Psi_{8}%
^{-}\left(  x\right)  }=\exp\left[  i2\arg\Psi_{8}^{+}\left(  x\right)
\right]  ,\text{ }x\in\left[  -1,-\infty\right)  ,
\end{equation}
where $\arg\Psi_{8}^{+}\left(  x\right)  =\operatorname{arccot}\left[
\operatorname{Re}\Psi_{8}^{+}\left(  x\right)  /\operatorname{Im}\Psi_{8}%
^{+}\left(  x\right)  \right]  $. Thus, the fundamental solution is
\begin{equation}
\varphi_{8}\left(  \zeta\right)  =e^{\gamma_{8}\left(  \zeta\right)  }\left(
\zeta+1\right)  ^{\lambda_{8}},
\end{equation}
where%
\begin{equation}
\gamma_{8}\left(  \zeta\right)  =\frac{1}{\pi}%
%TCIMACRO{\dint \nolimits_{-1}^{-\infty}}%
%BeginExpansion
{\displaystyle\int\nolimits_{-1}^{-\infty}}
%EndExpansion
dx\frac{\arg\Psi_{8}^{+}\left(  x\right)  }{x-\zeta}.
\end{equation}
The fundamental solution is chosen to satisfy Eq. (\ref{ch}). When choosing
$\arg\Psi_{8}^{+}\left(  -\infty\right)  =0$, $\arg\Psi_{8}^{+}\left(
-1\right)  =-2\pi$, the condition (\ref{II9})%
\begin{equation}
-\frac{1}{\pi}\arg\Psi_{8}^{+}\left(  -1\right)  +\lambda_{8}=0
\end{equation}
gives $\lambda_{8}=-2$. Therefore,%

\begin{equation}
\Psi_{8}\left(  \zeta\right)  =\eta_{8}\frac{\left(  \zeta-z\right)  \left(
\zeta-\omega\right)  }{\left(  \zeta+1\right)  ^{2}}e^{\gamma_{8}\left(
\zeta\right)  }. \label{zero8}%
\end{equation}
By setting $\zeta=0$, Eq. (\ref{zero8}) and its first- and second-order
derivatives give three equations for $z$, $\omega$, and the constant $\eta
_{8}$, whose solution is%
\begin{align}
z  &  =2\frac{P\lambda^{2}}{\left(  2j+1\right)  kT}\left\{  1+\frac
{P\lambda^{2}}{\left(  2j+1\right)  kT}\left[  \gamma_{8}^{\prime}\left(
0\right)  -2\right]  +\left\{  1-2\frac{P\lambda^{2}}{\left(  2j+1\right)
kT}\left[  \gamma_{8}^{\prime}\left(  0\right)  -\frac{3}{2}+2j\pi\frac
{a}{\lambda}\right]  \right.  \right. \nonumber\\
&  \left.  \left.  +\left(  \frac{P\lambda^{2}}{\left(  2j+1\right)
kT}\right)  ^{2}\left[  4\gamma_{8}^{\prime}\left(  0\right)  -\gamma
_{8}^{\prime}\left(  0\right)  ^{2}+2\gamma_{8}^{\prime\prime}\left(
0\right)  \right]  \right\}  ^{1/2}\right\}  ^{-1}. \label{2DFDPz}%
\end{align}
This relation between $z$\ and $T$\ is plotted in figure 8.

\begin{figure}[ptb]
\begin{center}
\includegraphics[height=2.5in]
{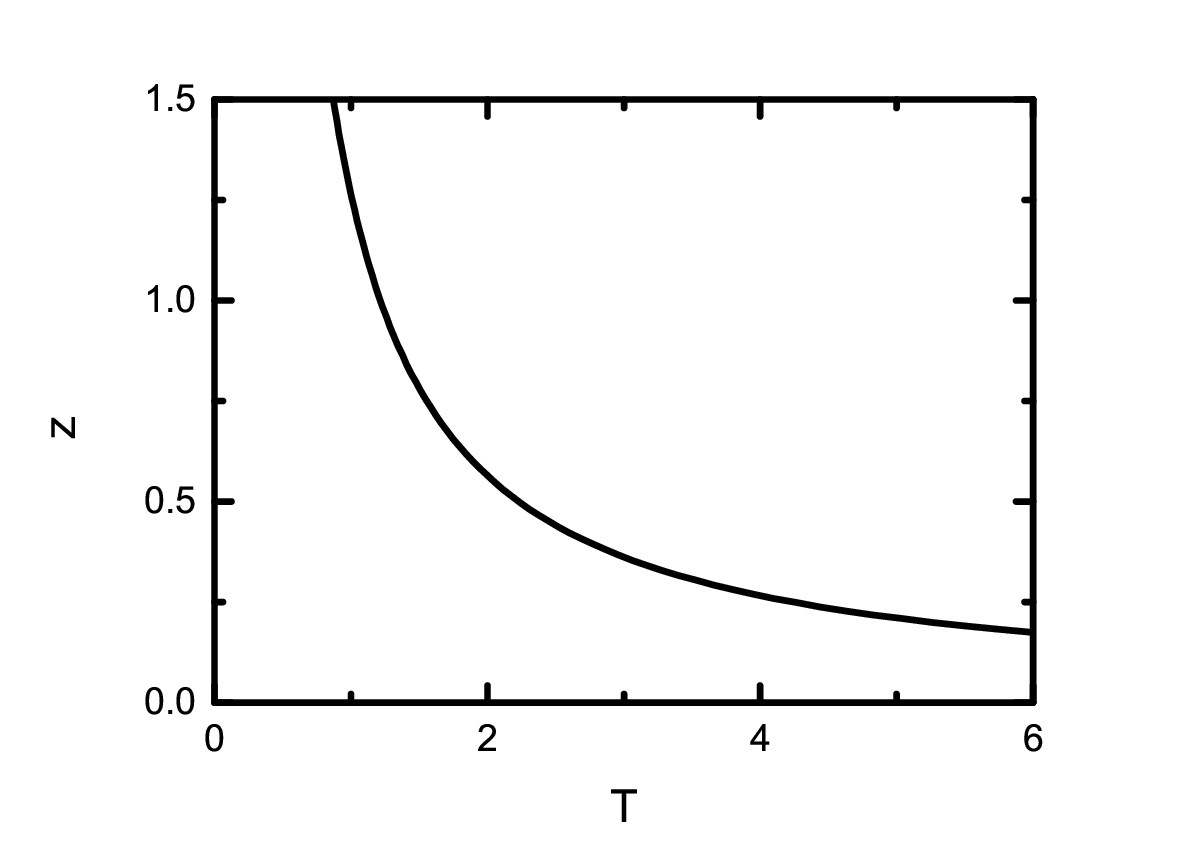}
\end{center}
\caption{The relation between $z$ and $T$ for two-dimensional hard-sphere
Fermi gases in an isobaric process.}%
\label{fig8}%
\end{figure}

Substituting the fugacity $z$ given by Eq. (\ref{2DFDPz}) into Eq.
(\ref{N2DF}) gives the equation of state.

\section{High temperatures and low densities \label{HTLD}}

The expression of the fugacity $z$ for hard-sphere quantum gases obtained in
the present paper is an exact result. In the conventional treatment, the
fugacity $z$ can only be obtained approximately. To compare with the equation
of state given by the conventional treatment, we expand the exact result
obtained above at high temperatures and low densities.

\subsection{Isochoric processes}

\subsubsection{Three-dimensional hard-sphere Bose gases}

The fugacity for three-dimensional hard-sphere Bose gases is given by Eq.
(\ref{3DBEVz}). At high temperatures and low densities, we can expand the
fugacity around $n\lambda^{3}/\left(  2j+1\right)  $ as%
\begin{equation}
z\simeq\frac{n\lambda^{3}}{2j+1}-\left[  \frac{1}{2\sqrt{2}}-4\left(
j+1\right)  \frac{a}{\lambda}\right]  \left(  \frac{n\lambda^{3}}%
{2j+1}\right)  ^{2}.
\end{equation}
Substituting it into Eq. (\ref{3DB1}) gives%
\begin{equation}
\frac{PV}{NkT}=1-\left[  \frac{1}{4\sqrt{2}}-2\left(  j+1\right)  \frac
{a}{\lambda}\right]  \frac{n\lambda^{3}}{2j+1}. \label{3DBEVeq}%
\end{equation}

\subsubsection{Three-dimensional hard-sphere Fermi gases}

At high temperatures and low densities, expanding the fugacity (\ref{3DFDVz})
as%
\begin{equation}
z\simeq\frac{n\lambda^{3}}{2j+1}+\left(  \frac{\sqrt{2}}{4}+4j\frac{a}%
{\lambda}\right)  \left(  \frac{n\lambda^{3}}{2j+1}\right)  ^{2}%
\end{equation}
and substituting it into Eq. (\ref{3DFDP}) give the equation of state%
\begin{equation}
\frac{PV}{NkT}=1+\left(  \frac{1}{4\sqrt{2}}+2j\frac{a}{\lambda}\right)
\frac{n\lambda^{3}}{2j+1}. \label{3DFDve}%
\end{equation}

\subsubsection{Two-dimensional hard-sphere Bose gases}

At high temperatures and low densities, expanding the fugacity (\ref{2DBEVz})
as%
\begin{equation}
z\simeq\frac{n\lambda^{2}}{2j+1}-\left[  \frac{1}{2}-2\pi\left(  j+1\right)
\frac{a}{\lambda}\right]  \left(  \frac{n\lambda^{2}}{2j+1}\right)  ^{2}%
\end{equation}
and substituting it into Eq. (\ref{2DBEP}) give the equation of state%
\begin{equation}
\frac{PV}{NkT}=1-\left[  \frac{1}{4}-\pi\left(  j+1\right)  \frac{a}{\lambda
}\right]  \frac{n\lambda^{2}}{2j+1}. \label{2DBEVeq}%
\end{equation}

\subsubsection{Two-dimensional hard-sphere Fermi gases}

At high temperatures and low densities, expanding the fugacity (\ref{2DFDVz})
as%
\begin{equation}
z\simeq\frac{n\lambda^{2}}{2j+1}+\left(  \frac{1}{2}+2\pi j\frac{a}{\lambda
}\right)  \left(  \frac{n\lambda^{2}}{2j+1}\right)  ^{2}%
\end{equation}
and substituting it into Eq. (\ref{2DFDP}) give the equation of state%
\begin{equation}
\frac{PV}{NkT}=1+\left(  \frac{1}{4}+\pi j\frac{a}{\lambda}\right)
\frac{n\lambda^{2}}{2j+1}. \label{2DFDve}%
\end{equation}

All the results given above agree with the result given by the virial
expansion \cite{JMP07}.

\subsection{Isobaric processes}

\subsubsection{Three-dimensional hard-sphere Bose gases}

In an isobaric process, there are two possible results of the fugacity $z$
given by Eqs. (\ref{ztwo0}) and (\ref{3DBEPz}), respectively.

(1) At high temperatures and low densities, expanding Eq. (\ref{ztwo0}) gives%

\begin{equation}
z\simeq\frac{P\lambda^{3}}{\left(  2j+1\right)  kT}-\left[  \frac{1}{4\sqrt
{2}}-2\left(  j+1\right)  \pi\frac{a}{\lambda}\right]  \left[  \frac
{P\lambda^{3}}{\left(  2j+1\right)  kT}\right]  ^{2}.
\end{equation}
Substituting it into Eq. (\ref{3DB2}) gives the equation of state,%
\begin{equation}
n=\frac{P}{kT}\left\{  1-\frac{1}{2j+1}\left[  -\frac{1}{4\sqrt{2}}+2\left(
j+1\right)  \frac{a}{\lambda}\right]  \frac{P\lambda^{3}}{kT}\right\}  .
\end{equation}
This result agrees with the virial expansion \cite{JMP07}.

(2) At high temperatures and low densities, expanding Eq. (\ref{3DBEPz}) gives%
\begin{equation}
z\simeq\frac{P\lambda^{3}}{\left(  2j+1\right)  kT}-\left[  \gamma_{5}%
{}^{\prime}\left(  0\right)  +1\right]  \left[  \frac{P\lambda^{3}}{\left(
2j+1\right)  kT}\right]  ^{2}.
\end{equation}
Substituting it into Eq. (\ref{3DB2}) gives the equation of state,%
\begin{equation}
n=\frac{P}{kT}\left\{  1-\frac{1}{\left(  2j+1\right)  }\left[  \gamma_{5}%
{}^{\prime}\left(  0\right)  +1-\frac{1}{2\sqrt{2}}+4\left(  j+1\right)
\frac{a}{\lambda}\right]  \frac{P\lambda^{3}}{kT}\right\}  .
\label{3DBGexpand}%
\end{equation}
The equation of state can be rewritten as the following form by solving $P$
from Eq. (\ref{3DBGexpand}):%
\begin{equation}
\frac{PV}{NkT}=1+\left[  \gamma_{5}{}^{\prime}\left(  0\right)  +1-\frac
{1}{2\sqrt{2}}+4\left(  j+1\right)  \frac{a}{\lambda}\right]  \frac
{n\lambda^{3}}{2j+1}.
\end{equation}

\subsubsection{Three-dimensional hard-sphere Fermi gases}

At high temperatures and low densities, expanding the fugacity (\ref{3DFDPz})
as%
\begin{equation}
z\simeq\frac{P\lambda^{3}}{\left(  2j+1\right)  kT}+\left(  \frac{1}{4\sqrt
{2}}+2j\frac{a}{\lambda}\right)  \left[  \frac{P\lambda^{3}}{\left(
2j+1\right)  kT}\right]  ^{2}%
\end{equation}
and substituting it into Eq. (\ref{3DFDn}) give the equation of state%
\begin{equation}
n=\frac{P}{kT}\left[  1-\frac{1}{2j+1}\left(  \frac{1}{4\sqrt{2}}+2j\frac
{a}{\lambda}\right)  \frac{P\lambda^{3}}{kT}\right]  .
\end{equation}

\subsubsection{Two-dimensional hard-sphere Bose gases}

At high temperatures and low densities, expanding the fugacity (\ref{2DBEPz})
as%
\begin{equation}
z\simeq\frac{P\lambda^{2}}{\left(  2j+1\right)  kT}-\left[  \frac{1}%
{4}-\left(  j+1\right)  \pi\frac{a}{\lambda}\right]  \left[  \frac
{P\lambda^{2}}{\left(  2j+1\right)  kT}\right]  ^{2}%
\end{equation}
and substituting it into Eq. (\ref{2DBEn}) give the equation of state%
\begin{equation}
n=\frac{P}{kT}\left\{  1-\frac{1}{2j+1}\left[  -\frac{1}{4}+\pi\left(
j+1\right)  \frac{a}{\lambda}\right]  \frac{P\lambda^{2}}{kT}\right\}  .
\end{equation}

\subsubsection{Two-dimensional hard-sphere Fermi gases}

At high temperatures and low densities, expanding the fugacity (\ref{2DFDPz})
as%
\begin{equation}
z\simeq\frac{P\lambda^{2}}{\left(  2j+1\right)  kT}+\left(  \frac{1}{4}%
+j\pi\frac{a}{\lambda}\right)  \left[  \frac{P\lambda^{2}}{\left(
2j+1\right)  kT}\right]  ^{2}%
\end{equation}
and substituting it into Eq. (\ref{N2DF}) give the equation of state%
\begin{equation}
n=\frac{P}{kT}\left[  1-\frac{1}{2j+1}\left(  \frac{1}{4}+j\pi\frac{a}%
{\lambda}\right)  \frac{P\lambda^{2}}{kT}\right]  .
\end{equation}

\section{Phase transitions \label{Phasetransitions}}

In this section, we discuss the problem of Bose-Einstein condensation phase
transition of three--dimensional hard-sphere Bose gases with the help of the
expression of $z$\ eq. (\ref{3DBEVz}).

According to eqs. (\ref{ztwo0}) and (\ref{3DBEPz}), we plot the relations
between $z$\ and $T$\ for different scattering length $a$\ in figure 9,
respectively. From the figure, we can see that each curve has an end, which is
the indication of a phase transition. As shown in figure 9, for $a<0$, the
transition temperature becomes lower with the increase of the interaction
strength, and the corresponding value of $z$\ is $1$. For $a>0$, the
transition temperature becomes higher with the increase of the interaction
strength, and the corresponding value of $z$ decreases.\ 

\begin{figure}[ptb]
\begin{center}
\includegraphics[height=2.5in]
{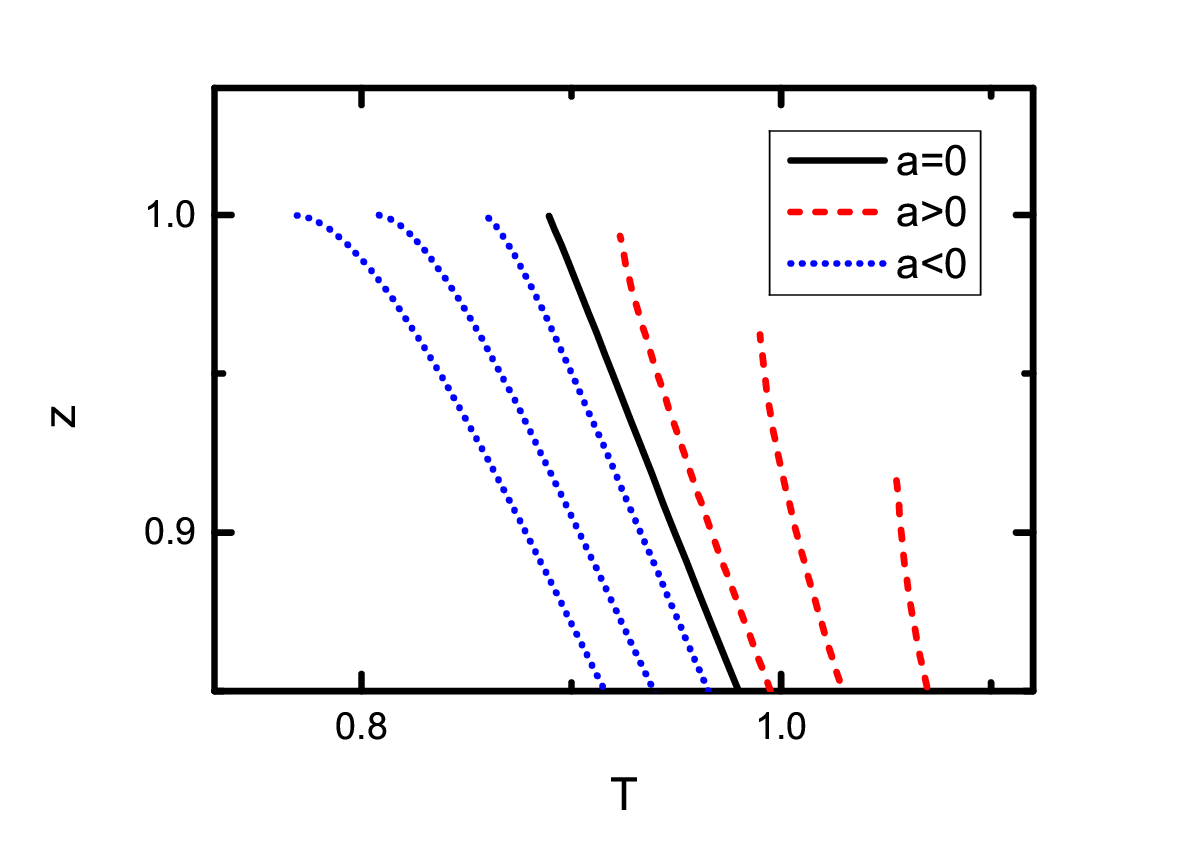}
\end{center}
\caption{The relation between $z$ and $T$ for three-dimensional hard-sphere
Bose gases in an isobaric process with different scattering length $a$. From
left to right $a$ is gradually increased from negative to positive.}%
\label{fig9}%
\end{figure}

\section{Conclusions \label{VI}}

To sum up, in this paper, we present the explicit expression of the fugacity
for two- and three-dimensional hard-sphere Bose and Fermi gases in isochoric
and isobaric processes. The method is to convert the problem of solving the
fugacity from the equation of state into the problem of finding the zero of a
complex function based on the homogeneous Riemann-Hilbert problem. This method
is introduced by Leonard for treating ideal quantum gases \cite{Leonard}.
Concretely, in this treatment, one can solve the fugacity $z$ from the
relation%
\begin{equation}
\Psi\left(  \zeta\right)  =\eta\left(  \zeta-z\right)
%TCIMACRO{\dprod \limits_{i=1}^{n_{\omega}-1}}%
%BeginExpansion
{\displaystyle\prod\limits_{i=1}^{n_{\omega}-1}}
%EndExpansion
\left(  \zeta-\omega_{i}\right)  \frac{\varphi\left(  \zeta\right)  }{%
%TCIMACRO{\dprod \limits_{k=1}^{n}}%
%BeginExpansion
{\displaystyle\prod\limits_{k=1}^{n}}
%EndExpansion
\left(  \zeta-c_{k}\right)  ^{\beta_{k}}},
\end{equation}
where $z$ and $\omega_{i}$, $i=1,\cdots,n_{\omega}-1$, are the zeros of the
complex function $\Psi\left(  \zeta\right)  $ which is constructed from the
equation of state of quantum gases, $n_{\omega}$ is the number of the zeros,
$c_{k}$ is the $k$-th end that is different from infinity of the boundary of
the analytic region of $\Psi\left(  \zeta\right)  $, $n$ is the number of the
ends different from infinity of the boundary, $\beta_{k}$ is a constant chosen
to ensure that $\varphi\left(  \zeta\right)  $ is a fundamental solution and
is determined by the behavior of $\Psi\left(  \zeta\right)  $ near the end
$c_{k}$, and $\varphi\left(  \zeta\right)  $ is a fundamental solution of the
homogeneous Riemann-Hilbert problem. By deriving both sides of this equation
at a given point or setting various values of $\zeta$, one can obtain a set of
equations of $z$, $\omega_{i}$, and the constant $\eta$. Solving such a set of
equations gives the fugacity $z$.

The key steps in this treatment are (1) analytically continuing the real
function $\Psi\left(  z\right)  $ whose zero corresponding to the fugacity $z$
to the whole complex plane, which gives the complex function $\Psi\left(
\zeta\right)  $, (2) finding the fundamental solution, and (3) determining the
number of zeros of $\Psi\left(  \zeta\right)  $. At the first step, in the
present case, the analytic continuation of $\Psi\left(  z\right)  $ relies on
the analytic continuation of Bose-Einstein and Fermi-Dirac integrals; the
analytically continued Bose-Einstein and Fermi-Dirac integrals are the
polylogarithm (Jonqui\'{e}re) functions, $Li_{\sigma}\left(  \zeta\right)  $
and $-Li_{\sigma}\left(  -\zeta\right)  $, respectively. At the second step,
based on the result of the analytic continuation, we can analyze the
singularity structure of $\Psi\left(  \zeta\right)  $, determine the boundary
of the analytic region of $\Psi\left(  \zeta\right)  $, calculate the jump of
$\Psi\left(  \zeta\right)  $ at the boundary, and then solve the fundamental
solution $\varphi\left(  \zeta\right)  $. At the third step, though in general
by the argument principle, one can only determine the difference between the
number of zeros and the number of isolated singularities, one can determine
the number of the zeros by use of the argument principle due to the fact that
$\Psi\left(  \zeta\right)  $ has no isolated singularities in the present case.

It is worthy to point out that the explicit expression of the fugacity for the
Bose case given in the present paper, especially the result for isobaric
processes, can be directly applied to analyze the phase transition of the
hard-sphere Bose gas system, which is an important problem in recent times
\cite{hsb,hsb2}. The reason is that the fugacity, and therefore the chemical
potential, plays a central role in the theory of phase transitions, according
to Erenfest's theory of phase transitions. In the Erenfest classification, the
order of a phase transition is defined by the order of the discontinuities in
the derivatives of the Gibbs free energy, $G=N\mu$. Our results show that the
transition temperature of the Bose-Einstein condensation can be obtained from
the relations between the fugacity $z$ and the temperature $T$.

\section*{Acknowledgement}

We are very indebted to Dr G. Zeitrauman for his encouragement. This work is
supported in part by NSF of China, under Project No. 11575125 and No. 11675119.

\end{document}